\documentclass[aps, twocolumn,showkeys,showpacs,amsmath,amssymb,superscriptaddress]{revtex4}
\usepackage{graphicx} 
\usepackage{dcolumn} 
\usepackage{bm}
\usepackage{epsfig}
\usepackage{subfigure}

\begin{document}
\title{Derivation of the Lattice Boltzmann Model for Relativistic
  Hydrodynamics}

\author{M. Mendoza} \email{mmendoza@ethz.ch} \affiliation{ ETH
  Z\"urich, Computational Physics for Engineering Materials, Institute
  for Building Materials, Schafmattstrasse 6, HIF, CH-8093 Z\"urich
  (Switzerland)}

\author{B. M. Boghosian}\email{bruce.boghosian@tufts.edu}
\affiliation{Bromfield-Pearson, Medford, Massachusetts 02155,
  Department of Mathematics, Tufts University }

\author{H. J. Herrmann}\email{hjherrmann@ethz.ch} \affiliation{ ETH
  Z\"urich, Computational Physics for Engineering Materials, Institute
  for Building Materials, Schafmattstrasse 6, HIF, CH-8093 Z\"urich
  (Switzerland)}

\author{S. Succi} \email{sauro.succi@gmail.com} \affiliation{Istituto
  per le Applicazioni del Calcolo C.N.R., Via dei Taurini, 19 00185,
  Rome (Italy),\\and Freiburg Institute for Advanced Studies,
  Albertstrasse, 19, D-79104, Freiburg, Germany}

\date{\today}
\begin{abstract}
  A detailed derivation of the Lattice Boltzmann (LB) scheme for
  relativistic fluids recently proposed in Ref.~\cite{rlbPRL}, is
  presented.  The method is numerically validated and applied to the
  case of two quite different relativistic fluid dynamic problems,
  namely shock-wave propagation in quark-gluon plasmas and the impact
  of a supernova blast-wave on massive interstellar clouds. Close to
  second order convergence with the grid resolution, as well as
  linear dependence of computational time on the number of grid points
  and time-steps, are reported.
\end{abstract}

\pacs{47.11.-j, 12.38.Mh, 47.75.+f}

\keywords{Lattice Boltzmann, quark-gluon plasma, relativistic fluid
  dynamics, supernovas}

\maketitle

\section{Introduction}

Relativistic fluid dynamics plays a major role in many fields of
modern physics, e.g. astrophysics, nuclear and high-energy physics
and, lately, also in condensed matter. The dynamics of such systems
requires solving highly nonlinear equations, rendering the analytic
treatment of practical problems extremely difficult. Therefore,
several numerical methods have been developed, based on macroscopic
continuum description \cite{NumMethod2, NumMethod4, NumMethod5} and
kinetic theory\cite{NumMethod1}.  Very recently, a new Lattice
Boltzmann (LB) scheme for relativistic fluids has been proposed, and
numerically validated for two rather different relativistic
applications, shock waves in quark-gluon plasmas and blast waves from
supernova explosions impinging against dense interstellar clouds
\cite{rlbPRL}.  This fills a missing entry in the remarkably
broad spectrum of LB applications across most areas of
fluid-dynamics, including quantum fluids \cite{QLB}.  While a
quantitative assessment of its practical impact on relativistic fluid
dynamics must necessarily await for a long and thorough validation
activity, work in Ref.~\cite{rlbPRL} provides robust indications that
the relativistic LB (RLB) stands concrete chances of carrying the
recognized advantages of LB schemes for classical fluids, over to the
relativistic context.  We refer primarily to mathematical
simplicity/computational efficiency, especially on 
parallel computers \cite{AMATI}, and easy handling of complex geometries.

In this paper, we present an extended version of our previous work
\cite{rlbPRL}.  First, we provide full details of the analytical and
numerical formulation leading to the relativistic LB scheme, including
the asymptotic (Chapman-Enskog) analysis of the continuum
fluid-dynamic limit, starting from the kinetic level.  The numerical
validation for the case of quark-gluon plasmas is also extended in
such a way as to probe the convergence/accuracy of the RLB schemes as
a function of grid resolution.  
Like for classical fluids, second-order convergence and linear
scaling of CPU time with number of grid points and time-steps is found.  
Moreover, the application to supernova
blast waves is explored in more detail, by investigating the effect of
increasing Lorentz factors on the space-time distribution of the
density and pressure fields.  Here, the numerical simulation of the
relativistic flow past a dense inter-stellar medium (massive clouds)
provides a clear indication that sweeping of interstellar matter
across the cloud becomes appreciable only for relativistic beta
factors $\beta > 0.5$.


\section{The basic idea}

The procedure developed \cite{rlbPRL} was prompted by two simple
observations, i) the kinetic formalism is naturally
covariant/hyperbolic, ii) being based on a finite-velocity, discrete
(beam) representation of the kinetic distribution function, standard
lattice Boltzmann methods naturally feature relativistic-like
equations of state, whereby the sound speed, $c_s$ is a sizeable
fraction of the speed of light $c$, i.e. the maximum velocity of mass
transport ($c_s/c=K$, with $0.1<K<1$).  Based on the above, and
choosing the lattice speed close to the value of the actual
light-speed (at variance with standard LB applications)
$c_l$$\equiv$$\delta x /\delta t \sim c$, the LB mathematical
framework allows the relativistic extension developed in our previous
work.  The standard LB reads as follows:
\begin{equation}
  \label{LBE}
  f_i(\vec{x} + \vec{c}_i \delta t;t+\delta t) - f_i(\vec{x};t) =-\omega \delta t \;(f_i-f_i^{eq}) \quad ,
\end{equation}
where $f_i(\vec{x};t)$ denotes the probability of finding a particle
at lattice site $\vec{x}$ and time $t$, moving along the direction
pointed by the discrete velocity $\vec{c}_i$.  The left-hand side is
readily recognized as an {\it exact} lattice transcription of the
free-streaming term $(\partial_t + v_a \nabla_a) f$ of the continuum
Boltzmann equation, where Latin index $a$ labels the spatial coordinates 
and repeated indices are summed upon.  
Being naturally covariant, this term goes virtually
unchanged over to the relativistic context.

The right-hand-side, on the other hand, is a discrete version of the
collision operator, here taking the form of a simple relaxation around
a local equilibrium $f_i^{eq}$, on a timescale $\tau = 1/\omega$.  The
local equilibrium encodes the symmetries/conservation laws governing
the ideal (non-dissipative) fluid regime, namely mass-momentum-energy
conservation and Galilean invariance.  While molecular details of the
collisional processes can be safely foregone, these conservation
properties must necessarily be preserved in the lattice formulation.

The discrete local equilibrium is usually expressed as a local
Maxwellian, expanded to second order in the local Mach number
$Ma=u/c_s$, $u$ being the local flow speed.  For the case of athermal
flows, this takes the form
\begin{equation}
  f_i^{eq} = w_i \rho \left(1 + \frac{c_{ia} u_a}{c_s^2} + \frac{Q_{iab} u_a u_b}{2 c_s^4} \right) \quad ,
\end{equation}
where (particle mass is taken to unity for simplicity):
\begin{eqnarray*}
  \begin{aligned}
    \rho &= \sum_i f_i \quad , \\
    \rho u_a &= \sum_i f_i c_{ia} \quad ,
  \end{aligned}
\end{eqnarray*}
are the fluid density and mass current density, respectively.  In the
above, $w_i$ is a set of weights, obeying the sum-rules $\sum_i w_i
=1$ and $\sum_i w_i c_{ia}^2 = c_s^2$, and $Q_{iab} = c_{ia}
c_{ib}-c_s^2 \delta_{ab}$ is the projector along the $i$-th spatial
direction.  It is readily checked that the local equilibria fulfill
the following conservation rules
\begin{eqnarray}
\label{CON}
\begin{aligned}
  \sum_i f_i^{eq} &= \sum_i f_i = \rho \quad ,\\
  \sum_i f_i^{eq} c_{ia} &= \sum_i f_i c_{ia} = \rho u_a \quad ,\\
  \sum_i f_i^{eq} c_{ia} c_{ib} &= \rho (u_a u_b + c_s^2 \delta_{ab})
  \quad .
\end{aligned}
\end{eqnarray}
The first two are the usual mass-momentum conservations laws, whereas
the latter ensures the isotropy of the equilibrium momentum-flux.  
The latter is crucial to secure the proper non-linear structure of the
Navier-Stokes equations, and indeed only specific classes of discrete
lattices fulfill the aforementioned conservation constraints.  As
previously noted, lattice equilibria can be obtained by local
expansion of the continuum expression of local Maxwell equilibria.  In
a more empirical way, they could also be obtained by matching the
local equilibria in parametric form, $A e^{-B c_{ia} u_a}$, to the
conservation rules \cite{CON}, thereby fixing the Lagrangian
parameters $A$ and $B$ in terms of the conserved hydrodynamic fields
$\rho$ and $u_a$.  The possibility of fixing local equilibria by
simply expanding the local continuum Maxwellian, which is more elegant
than empirical matching \cite{HERMI}, is by no means evident.  

In fact, it is strictly related to the well-known property of the local
Maxwellian to serve as the generating function of Hermite's polynomials
$H_n$ (here $v =v/c_s$ and $u=u/c_s$);
\begin{equation}
\label{GENFUN}
e^{-\frac{(v-u)^2}{2}} = e^{-\frac{v^2}{2}} \sum_{n=0}^{\infty} H_n(v) u^n \quad .
\end{equation}

Note that the Galilean invariance manifestly encoded at the right-hand
side through the dependence on the magnitude of
the relative speed $(v_a-u_a)$, can only be
preserved by including {\it all} terms in the Mach-number expansion at
the right hand side.  It is quite fortunate that the Navier-Stokes
equations only involve quadratic non-linearities in the flow field,
because this allows to develop a consistent lattice hydrodynamic
theory by retaining only second order terms in the Mach-number
expansion.  A similar line of thinking can also be applied to the
relativistic equations, with due changes in the mathematical-physical
details, to be discussed shortly.

On the other hand, we are not aware of any relativistic analogue of
the relation (\ref{GENFUN}) for relativistic local equilibria
(J\"uttner distribution).  Because of this, the relativistic LB scheme
has been devised according to the moment-matching procedure discussed
above.  That is, the local kinetic equilibria are expressed as
parametric polynomials of the relativistic fluid velocity $\vec{\beta}
= \vec{u}/c$, with the Lagrangian parameters fixed by the condition of
matching the analytic expression of the relevant relativistic moments,
namely the number density, energy density and energy-momentum.  As
anticipated, the possibility of a successful matching stems directly
from the fact that, even in standard (non-relativistic) LB fluids, the
sound speed $c_s$ is of the same order of the speed of light,
typically $c_s=c/\sqrt 3$, which is exactly the equation of state of
ideal relativistic fluids.  As a result, $|\vec{\beta}| = Ma/\sqrt 3$,
so that $|\vec{\beta}|$ is of the same order as the Mach number
$Ma=|\vec{u}|/c_s$.  Thanks to this simple, and yet basic property, it
is possible to tackle weakly relativistic problems in close analogy
with the LB theory of classical low-Mach fluids, the algebraic details
being of course quite different in the two cases.

This permits to carry most of the LB formalism over to the context of
weakly relativistic fluids, such as quark-gluon plasmas generated by
recent experiments on heavy-ions and hadron jets \cite{QG-1, QG-2,
  QG-3, QG-4, QG-5, QG-6, QG-7}, as well as astrophysical flows, such
as interstellar gas and supernova remnants \cite{Scipaper, supernova1,
  supernova2, supernova3}.

The RLB scheme is verified through quantitative comparison with recent
one dimensional hydrodynamic simulations of relativistic shock wave
propagation in viscous quark-gluon plasmas \cite{BAMPSs}, and also
applied to the three dimensional case of a blast-wave, produced by a
supernova explosion, colliding against interstellar massive matter,
e.g. molecular gas \cite{Scipaper}.

Being based on a second-order moment-matching procedure, rather than
on a high-order systematic expansion in $\vec{\beta}$ of the local
relativistic equilibrium (J\"uttner) distribution\cite{NumMethod1},
the RLB is limited to weakly relativistic problems, with
$|\vec{\beta}| \sim 0.1$.  Note in fact that, unlike the continuum
Maxwellian, polynomial expansions are positive-definite only for
Mach-number (relativistic $\beta$) below a given threshold, typically
$Ma \sim 0.3$.
However, by introducing artificial faster-than-light particles
(numerical ``tachyons''), the RLB scheme can be taken up to
$|\vec{\beta}| \sim 0.6$, corresponding to Lorentz's factors $\gamma
=\frac{1}{\sqrt{1-|\vec{\beta}|^2}} \sim 1.4$ \cite{rlbPRL}. Although
still far from strongly relativistic regimes, with $\gamma \gg 10$ and
higher, this Lorentz factor is nevertheless relevant to a host of
important relativistic fluid problems at wildly disparate scales, such
as quark-gluon plasmas and relativistic outflows in supernova
explosions and possibly even Dirac fluids in graphene \cite{GRAPHENE}.


\section{Model Description}

We begin our model description by considering the relativistic fluid
equations associated with the conservation of number of particles and
momentum-energy. The energy-momentum tensor reads as
follows\cite{RelaBoltEqua, paperRomat}: $T^{\mu \nu}= P \eta^{\mu \nu}
+ (\epsilon +P)u^\mu u^\nu + \pi^{\mu \nu}$, $\epsilon$ being the
energy density, $P$ the hydrostatic pressure and $\pi^{\mu \nu}$ the
dissipative component of the stress-energy tensor, to be specified
later. The velocity 4-vector is defined by $u^{\mu}= (\gamma, \gamma
\vec{\beta})^{\mu}$, where $\vec{\beta}=\vec{u}/c$ is the velocity of
the fluid in units of the speed of light and
$\gamma$$=$$\frac{1}{\sqrt{1-|\vec{\beta}|^2}}$. The tensor $\eta^{\mu
  \nu}$ denotes the Minkowski metric.  Additionally, we define the
particle 4-flow, $N^{\mu}= n \gamma (1, \vec{\beta})^{\mu}$,
with $n$ the number of particles per volume.  Applying the
conservation rule to energy and momentum, $\partial_\mu T^{\mu \nu} =
0$, and to the 4-flow, $\partial_\mu N^{\mu} = 0$, we obtain the
hydrodynamic equations,
\begin{subequations}\label{macroeq10}
  \begin{eqnarray}
    \begin{aligned}
      \partial_t\left( (\epsilon+P)\gamma^2-P \right)
      &+ \partial_a \left((\epsilon + P)\gamma^2 u_a \right) \\
      &+\partial_t \pi^{00} + \partial_a \pi^{a 0} = 0 \quad ,
    \end{aligned}
  \end{eqnarray}
  \begin{eqnarray}
    \begin{aligned}
      \partial_t\left( (\epsilon+P)\gamma^2 u_b \right) + \partial_b P
      &+ \partial_a \left( (\epsilon + P)\gamma^2 u_a u_b \right) \\
      &+ \partial_t \pi^{0 b} + \partial_a \pi^{a b} = 0 \quad ,
    \end{aligned}
  \end{eqnarray}
\end{subequations}
for the energy momentum conservation, and
\begin{equation}\label{macroeq20}
  \partial_t (n \gamma) + \partial_a \left( n\gamma u_a \right) = 0 \quad ,
\end{equation}
for the conservation of particle number. Note that, unlike the case of
non-relativistic fluids, we have two scalar equations, one for
the particle number and one for the energy (in classical hydrodynamics
energy appears as the trace of a second-order moment, namely the
momentum-flux) .  To complete the set of equations, we need to define
a state equation relating at least two of the three quantities: $n$,
$P$ and $\epsilon$.

\subsection{Relativistic Boltzmann Equation}
The above hydrodynamic equations can be derived as a macroscopic limit
of the relativistic Boltzmann equation. 
For the case of a single non-degenerated gas, and in the absence of 
external forces, this reads as follows\cite{RelaBoltEqua}:
\begin{equation}\label{RBE}
  \partial_{\mu} (p^{\mu} f) = \int(f'_{*}f' - f_{*}f) 
  \Phi \sigma d\Omega \frac{d^3p_{*}}{p_{* 0}} \quad , 
\end{equation}
where $p^{\mu} = \left (\frac{E(p)}{c}, \vec{p}\right )$ is the
particle 4-momentum with $E(p)$ the relativistic energy as function of
the momentum magnitude $p$$=$$|\vec{p}|$, $E(p)=(p^2 c^2 + m^2
c^4)^{1/2}$.  In the above, $f_{*}$$\equiv$$f(\vec{x},\vec{p}_{*},t)$
and $f$$\equiv$$f(\vec{x},\vec{p},t)$ denote the distribution
functions before the collision, while
$f'_{*}$$\equiv$$f(\vec{x},\vec{p}'_{*},t)$ and
$f'$$\equiv$$f(\vec{x},\vec{p}',t)$ are the resulting ones after the
collision. The base of the so-called collision cylinder is described
by $\sigma d\Omega$, with $\sigma$ the differential cross section,
$\Omega$ is the solid angle, and
\begin{equation}
  \begin{aligned}
    \Phi &= \frac{p^0 p^0_*}{c}\sqrt{(\vec{v}-\vec{v}_*)^2 -
      \frac{1}{c^2}(\vec{v}\times \vec{v}_*)^2 } \\ &= \sqrt{
      (p^{\mu}_* p_\mu)^2 - m^2 c^4}
  \end{aligned}
\end{equation}
is the Lorentz invariant flux \cite{RelaBoltEqua}, with $\vec{v}$ and
$\vec{v}_*$ the velocity of the particles with momentum $\vec{p}$ and
$\vec{p}_*$, respectively. The right-hand-side of Eq.~\eqref{RBE} is
the collision term, whose details fix the value of the transport
coefficients in the macroscopic equations.  Although the collision
integral can be expressed in terms of the second kind modified Bessel
functions and numerical integrations \cite{RelaBoltEqua}, simpler
expressions have been proposed, along the lines of the BGK
(Bhatnagar-Gross-Krook) approximation for non-relativistic fluids.
The first relativistic BGK (RBGK), as proposed by
Marle\cite{MarleModel}, reads as follows:
\begin{equation}
  \label{RBGKM}
  \partial_{\mu} (p^{\mu} f) = \frac{m}{\tau_M}(f^{eq}-f) \quad , 
\end{equation}
where $f^{eq}$ is a local relativistic equilibrium, $m$ is the
particle {\it rest} mass, and $\tau_M$ represents a characteristic
time between subsequent collisions. This can be regarded as the
relaxation time only in a local rest frame where the momentum of the
particles is zero\cite{RelaBoltEqua}. It is well-known that in a
general inertial frame, the relaxation time $\hat \tau_{M}$ can be
written as follows:
\begin{equation}
  \label{relaxtime}
  \hat \tau_{M} = \frac{p^0}{mc} \tau_M \quad . 
\end{equation}
Although, in the Marle model, the transport coefficients are expressed
usually as functions of the characteristic time $\tau_M$, they cannot
be described as functions of the relaxation time $\hat \tau_M$ because
it depends on the microscopic momentum component $p^0$, which means on
microscopic $\gamma_v$$=$$\frac{1}{\sqrt{1-\frac{|\vec{v}|^2}{c^2}}}$,
and therefore it cannot appear in any macroscopic description. To
avoid this problem, Takamoto and Inutsuka \cite{Takamoto} proposed a
modified Marle model, in which the relaxation time $\tau$ is taken as
the weighted average, i.e.  $\frac{1}{\tau}$$=$$\langle \frac{1}{\hat
  \tau_{M}} \rangle$. With this approximation, the following relation
can be obtained \cite{Takamoto, RelaBoltEqua}
\begin{equation}
  \label{relaxtime2}
  \tau_M = \frac{K_1(\chi)}{K_2(\chi)} \; \tau \quad , 
\end{equation}
where $\chi$$\equiv $$\frac{mc^2}{kT}$ and $K_n$ is the second kind
modified Bessel function of order $n$. The correction
$\frac{K_1(\chi)}{K_2(\chi)}$ tends to $1$ at low temperatures, i.e.
$\chi$$\rightarrow $$\infty$, and to $\frac{\chi}{2}$ in the limit of
high temperatures, i.e. $\chi$$\rightarrow $$0$. In this modified
approach, the characteristic time in the transport coefficient can be
replaced by the relaxation time $\tau$ using Eq.~\eqref{relaxtime2} as
an approximation.

The Marle model provides a good approximation of the full collision
term at low temperatures.

A more general RBGK model, which provides a reasonable approximation
of the transport coefficients at both low and high temperatures, was
subsequently proposed by Anderson and Witting \cite{Anderson}, and it
reads as follows:
\begin{equation}
\label{RBGKA}
\partial_{\mu} (p^{\mu} f) = \frac{u^{\mu}p_{\mu}}{c^2 \tau_A}(f^{eq}-f) \quad , 
\end{equation}
$\tau_A$ being the relaxation time. 

Both models can reproduce on the macroscopic level the conservation
equations given by $\partial_\mu T^{\mu \nu} = 0$, and $\partial_\mu
N^{\mu} = 0$, although with different expressions for the dissipative
terms and transport coefficients.  For instance, the shear viscosity
using the Marle model is given by $\eta_M$$\simeq$$\frac{4 P^{eq}
  \tau_M}{\chi}$ for high temperatures (ultra-relativistic case), with
$P^{eq}$ the equilibrium pressure, while with the Anderson-Witting
model yields $\eta_A$$\simeq$$\frac{4 P^{eq} \tau_A}{5}$.

In general, the dissipation parameters, like the bulk viscosity, thermal
conductivity and shear viscosity, are only approximations of the values 
obtained by linearization of the full collision term in the relativistic 
Boltzmann equation, Eq.\ref{RBE}.

Having discussed the BGK formulation in the relativistic context, we
next proceed to map it within the Lattice Boltzmann framework.

\subsection{Lattice Boltzmann Model}
\label{sec:lbmodel}
The Lattice Boltzmann theory for classical fluids shows that it may
prove more convenient to solve fluid problems by numerically
integrating the underlying kinetic equation rather than the
macroscopic fluid equation themselves. The main condition for this to
happen is that a sufficiently economic representation of the velocity
degrees of freedom be available.  Following upon consolidated
experience with non-relativistic fluids, such a representation is
indeed provided by discrete lattices, whereby the particle velocity
(momentum) is constrained to a handful of constant discrete
velocities, with sufficient symmetry to secure isotropy and the
fundamental conservations of fluid flows, namely mass-momentum-energy
conservation and rotational invariance.  The main advantages of the
kinetic representation of classical fluids have been discussed at
length\cite{LBE2}, and they amount basically to the fact that the
information is transported along straight-streamlines (the discrete
velocities are constant in space and time) rather than along
space-time dependent trajectories generated by the flow itself, as it
is case for hydrodynamic equations.  Moreover, diffusive transport is
not described by second-order spatial derivatives, but rather emerges
as a collective property from the adiabatic relaxation of the momentum
flux tensor to its local equilibrium value.  This is crucial in
securing a balance between first-order derivatives in both space and
time, which is essential for relativistic equations.

\begin{figure}
  \centering \includegraphics[scale=0.4]{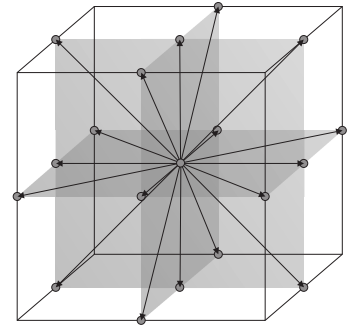}
  \caption{Set of discrete velocities for the relativistic lattice
    Boltzmann model. The highest speed is $\sqrt{2}c_l.$}\label{d3q19}
\end{figure}

In order to reproduce the relativistic hydrodynamic equations, an LB
model with the D3Q19 ($19$ speeds in $3$ spatial dimensions) cell
configuration, as shown in Fig.~\ref{d3q19}, was proposed in
Ref.~\cite{rlbPRL}.  From Fig.~\ref{d3q19} it is readily appreciated
that the highest D3Q19 speed is $\sqrt{2}c_l$, $c_l$$=$$\frac{\delta
  x}{\delta t}$ being the limiting lattice speed along each direction.
The velocity units are rescaled such that the speed of light becomes
$c$$=$$1$.

As noted above, relativistic hydrodynamics evolves two scalars, number
and energy density.  It is therefore convenient to introduce two
separate distribution functions $f_i$ and $g_i$ for each velocity
vector $\vec{c}_{i}$, representing, so to say, ``fluons'' and
``phonons'', respectively.

The hydrodynamic variables are calculated by using the following five
macroscopic constraints,
\begin{subequations}\label{macrosLB}
  \begin{equation}
    n\gamma=\sum_{i=0}^{18} f_{i} \quad ,
  \end{equation}
  \begin{equation}
    (\epsilon + P)\gamma^2 - P= \sum_{i=0}^{18} g_{i} \quad ,
  \end{equation}
  \begin{equation}
    (\epsilon + P)\gamma^2 \vec{u}=\sum_{i=0}^{18} g_{i} \vec{c}_{i} \quad ,
  \end{equation}
\end{subequations}

From these equations, we need to extract six physical quantities,
namely $n$, $\vec{u}$, $\epsilon$ and $P$.  With five equations for
six unknowns, the problem is closed by choosing an equation of state,
which we take of the form $\epsilon$$=$$3P$ \cite{RelaBoltEqua}.  We
wish to emphasize that the present LB scheme is by no means limited to
this choice.

Both distribution functions $f_i$ and $g_i$ are postulated to evolve
according to the relativistic Boltzmann-BGK equation {\it based on the
low-temperature Marle model}, Eq.~\eqref{RBGKM}. 

To obtain the lattice analogue of the Marle model, we first
write explicitly Eq.~\eqref{RBGKM} as follows:
\begin{equation}\label{RBGK1}
  \partial_0 (p^0 f) + \partial_a (p^a f) = \frac{m}{\tau_M}(f^{\rm eq}-f) \quad .
\end{equation}
Replacing the value of the four-momentum, we obtain
\begin{equation}\label{RBGK2}
  \partial_0 (m \gamma_v f) + \partial_a ( m c^a \gamma_v f) = \frac{m}{\tau_M} (f^{\rm eq}-f) \quad ,
\end{equation}
with $\gamma_v$ the Lorentz factor for the microscopic velocities
$c^a$. Due to the fact that the velocity and spatial coordinates are
linearly independent, we can further write:
\begin{equation}\label{RBGK3}
  \gamma_v \partial_0 f +  \gamma_v c^a\partial_a f = \frac{f^{\rm eq}-f}{\tau_M} \quad .
\end{equation}
Dividing by $\gamma_v$ on both sides of Eq.~\eqref{RBGK3}, we obtain
\begin{equation}\label{RBGK4}
  \partial_0 f +  c^a\partial_a f = \frac{mc}{\tau_M p^0}(f^{\rm eq}-f) \quad ,
\end{equation}
and replacing Eq.~\eqref{relaxtime}, we obtain
\begin{equation}\label{RBGK5}
\partial_0 f + c^a\partial_a f = \frac{1}{\hat \tau_{M}}(f^{\rm eq}-f) \quad .
\end{equation}
According to the modified Marle model \cite{Takamoto}, we can write
Eq.~\eqref{RBGK5} as
\begin{eqnarray}\label{RBGK6}
  \begin{aligned}
    \partial_0 f + c^a\partial_a f = \left (\frac{1}{\tau}-\vartheta
    \right) \;(f^{\rm eq}-f) ,
  \end{aligned}
\end{eqnarray}
where the correction term $\vartheta$, using Eq.~\eqref{relaxtime2},
is given by
\begin{eqnarray}\label{RBGK7}
  \begin{aligned}
    \vartheta &= \left[ \left \langle \frac{1}{\tau_{M*}} \right
      \rangle - \frac{1}{\tau_{M*}} \right]\\& =
    \frac{1}{\tau_M}\left[ \frac{K_1(\chi)}{K_2(\chi)} -
      \frac{1}{\gamma_v} \right]\quad .
  \end{aligned}
\end{eqnarray}
For low temperatures, $\frac{K_1(\chi)}{K_2(\chi)} \sim 1$ and
$\gamma_v \sim 1$, so that the correction term $\vartheta$ tends to
zero, thereby renstituting the non-relativistic Boltzmann equation.
At high temperatures, this term can be approximated by
\begin{eqnarray}\label{RBGK7}
  \begin{aligned}
    \vartheta \sim \frac{\chi}{2} \frac{1}{\tau_M}\quad,
  \end{aligned}
\end{eqnarray}
which also tends to vanish as temperature is made higher.

As noted above, Eq.~\eqref{RBGK6}, without the term $\vartheta$, is
just the Boltzmann equation for the case of non-relativistic fluids,
with the collision time $\tau$ representing a realistic relaxation
time of the system.

Therefore, for the purpose of this work, we postulate the discrete
distribution functions to evolve according to the following pair of
BGK Boltzmann equations \cite{BGK},
\begin{eqnarray}{\label{lbe1}}
  f_{i}(\vec{x}+\vec{c}_i \delta t,t+\delta
  t)-f_{i}(\vec{x},t)&=-\frac{\delta t}{\tau} (f_i - f_i^{\rm eq}) \quad ,
\end{eqnarray} 
and,
\begin{eqnarray}{\label{lbe2}}
  g_{i}(\vec{x}+\vec{c}_i\delta t,t+\delta
  t)-g_{i}(\vec{x},t)&=-\frac{\delta t}{\tau} (g_i - g_i^{\rm eq}) \quad ,
\end{eqnarray}
where $f_i^{\rm eq}$ and $g_i^{\rm eq}$ are the equilibrium
distribution functions.

To find the equilibrium distribution functions recovering the
relativistic fluid equations, Eqs.~\eqref{macroeq10} and
\eqref{macroeq20}, in the continuum limit, we use the moment-matching
procedure described earlier on in this paper.

More precisely, we write the equilibrium distribution functions as,
\begin{subequations}\label{equil0}
  \begin{equation}
    f_i^{\rm eq} = w_i [A + \vec{c}_i\cdot \vec{B}] \quad , {\rm for} \quad i \geq 0 \quad ,
  \end{equation}
  \begin{equation}
    g_i^{\rm eq} = w_i [C + \vec{c}_i\cdot \vec{D} + \tensor{E}:(\vec{c}_i \vec{c_i} - \alpha  \tensor{I})] \quad , {\rm for} \quad i > 0 \quad ,
  \end{equation}
  \begin{equation}
    g_0^{\rm eq} = w_0 [F] \quad , 
  \end{equation}
\end{subequations}
where $\alpha$, $A$, $\vec{B}$, $C$, $\vec{D}$, $\tensor{E}$ and $F$
are Lagrange parameters, to be fixed by matching the discrete to the
correct continuum equations. The weights $w_i$ for this set of
discrete speeds are defined by $w_{0}=1/3$ for the rest particles,
$w_i=1/18$ for the velocities $|\vec{c}_i|$$=$$c_l$, and $w_i=1/36$
for $|\vec{c}_i|$$=$$\sqrt{2}c_l$.

First, we find the values for $A$ and $\vec{B}$ to obtain the
conservation of particle number, Eq.~\eqref{macroeq20}.  To this
purpose, we impose
\begin{equation}
  \sum_{i=0}^{18} f_i^{\rm eq} = n \gamma \quad ,
\end{equation}
and,
\begin{equation}
  \sum_{i=0}^{18} f_i^{\rm eq} \vec{c}_i = n \gamma \vec{u} \quad .
\end{equation}
Replacing the Eq.~\eqref{equil0} into the sums, we arrive to
\begin{equation}
  \sum_{i=0}^{18} f_i^{\rm eq} = A = n \gamma \quad ,
\end{equation}
and
\begin{equation}
  \sum_{i=0}^{18} f_i^{\rm eq} \vec{c}_i = \frac{3}{c_l^2}\vec{B} = n \gamma \vec{u} \quad ,
\end{equation}
where, we can see easily that $A$$=$$n\gamma$ and
$\vec{B}$$=$$\frac{3}{c_l^2}\gamma n \vec{u}$.  Next, we have to
obtain the Eq.~\eqref{macroeq10} from the equilibrium distribution
functions $g_i^{\rm eq}$. To this end, we impose the following
constraints:
\begin{equation}
  \sum_{i=0}^{18} g_i^{\rm eq} = \gamma^2 (\epsilon + P) - P \quad ,
\end{equation}
\begin{equation}
  \sum_{i=0}^{18} g_i^{\rm eq} \vec{c}_i = (\epsilon + P) \gamma^2 \vec{u} \quad .
\end{equation}
and additionally,
\begin{equation}
  \sum_{i=0}^{18} g_i^{\rm eq} c_{ia} c_{i\beta} = P\delta_{a b} + (\epsilon + P)\gamma^2 u_a u_b \quad .
\end{equation}
Using a similar procedure as before, we can find the rest of the
Lagrange parameters, $\alpha$$=$$\frac{c_l^2}{3}$,
$C$$=$$\frac{3P}{c_l^2}$, $\vec{D}$$=$$\frac{3}{c_l^2}(\epsilon +
P)\gamma^2 \vec{u}$, $E_{a b}$$=$$\frac{9}{2c_l^4}(\epsilon +
P)\gamma^2u_a u_b$, and $F$$=$$(\epsilon + P)\gamma^2 \left[3 -
  3\frac{(2+c_l^2)P}{c_l^2(\epsilon + P)\gamma^2} -
  \frac{3}{2c_l^2}(\epsilon + P)\gamma^2 |\vec{u}|^2 \right]$.  These
calculations are shown in detail in Appendix \ref{LagrangeParameters}.

The equilibrium distribution functions recovering the relativistic
fluid equations in the continuum limit, finally read as follows:
\begin{eqnarray}{\label{equil1}}
  f_i^{\rm eq} = w_i n \gamma \left[1+3\frac{(\vec{c}_i \cdot \vec{u})}{c_l^2} \right] \quad ,
\end{eqnarray} 
for $i$$\ge$$0$,
\begin{eqnarray}{\label{equil2a}}
  \begin{aligned}
    g_i^{\rm eq} = w_i (\epsilon +P) \gamma^2 \biggl[\frac{3
      P}{(P+\epsilon)\gamma^2 c_l^2} +3\frac{(\vec{c}_i \cdot
      \vec{u})}{c_l^2} \\ + \frac{9}{2}\frac{(\vec{c}_i \cdot
      \vec{u})^2}{c_l^4} - \frac{3}{2}\frac{|\vec{u}|^2}{c_l^2}
    \biggr] \quad ,
  \end{aligned}
\end{eqnarray} 
for $i$$>$$0$, and
\begin{eqnarray}{\label{equil2b}}
  g_0^{\rm eq} = w_0 (\epsilon +P) \gamma^2 \left[ 3 - \frac{3 P
      (2+c_l^2)}{(P+\epsilon)\gamma^2 c_l^2} -
    \frac{3}{2}\frac{|\vec{u}|^2}{c_l^2} \right] \quad ,
\end{eqnarray} 
for the rest particles. 

By Taylor expanding the Eqs. \eqref{lbe1} and \eqref{lbe2} to second
order in $\delta t$, and retaining terms only up to first order in the
Chapman-Enskog expansion $f=f^{eq}+ \kappa f^{1}+\dots$, where $\kappa \sim
c \tau \nabla$ is the Knudsen number, the LB equations can be shown to
reproduce the following continuum fluid equations as derived in detail
in Appendix \ref{ChapmanEnskog}:
\begin{subequations}\label{macroeq1}
  \begin{equation}\label{macroeq1a}
    \partial_t\left[ (\epsilon+P)\gamma^2-P \right] + \partial_a
    \left[(\epsilon + P)\gamma^2 u_a \right] = 0 \quad ,
  \end{equation}
  \begin{equation}\label{macroeq1b}
    \begin{aligned}
      \partial_t & \left[ (\epsilon+P)\gamma^2 u_b \right] +
      \partial_b
      P + \partial_a \left[ (\epsilon + P)\gamma^2 u_a u_b \right] \\
      &= \partial_a \left[ \partial_b (\eta \gamma u_a)+
        \partial_a(\eta \gamma u_b) + \partial_l (\eta \gamma
        u_l)\delta_{ab} \right] ,
    \end{aligned}
  \end{equation}
\end{subequations}
for the energy momentum conservation, and
\begin{equation}\label{macroeq2}
  \partial_t (n \gamma) + \partial_a \left( n\gamma u_a\right) = 0 \quad ,
\end{equation}
for the conservation of particle number.  The indices $a$,$b$ and $l$
denote the spatial components.

The choice of the state equation, $\epsilon$$=$$3P$, simplifies the
equilibrium functions as follows,
\begin{eqnarray}{\label{equil1s}}
  f_i^{\rm eq} = w_i n \gamma \left[1+3\frac{(\vec{c}_i \cdot \vec{u})}{c_l^2} \right] \quad ,
\end{eqnarray} 
for $i$$\ge$$0$ and,
\begin{eqnarray}{\label{equil2as}}
  \begin{aligned}
    g_i^{\rm eq} = w_i \epsilon \gamma^2 \biggl[ \frac{1}{\gamma^2
      c_l^2} +4\frac{(\vec{c}_i \cdot \vec{u})}{c_l^2}+
    6\frac{(\vec{c}_i \cdot \vec{u})^2}{c_l^4} -
    2\frac{|\vec{u}|^2}{c_l^2} \biggr] \quad ,
  \end{aligned}
\end{eqnarray} 
for $i$$>$$0$ and,
\begin{eqnarray}{\label{equil2bs}}
  g_0^{\rm eq} = w_0 \epsilon \gamma^2 \left[ 4 - \frac{2+c_l^2}{\gamma^2
      c_l^2} - 2\frac{|\vec{u}|^2}{c_l^2} \right] \quad ,
\end{eqnarray} 
for $i$$=$$0$. Then, the equations for the macroscopic variables take
the form: $n\gamma=\sum_{i=0}^{18} f_{i}$, $\frac{4}{3}\epsilon
\left(\gamma^2 - \frac{1}{4} \right)= \sum_{i=0}^{18} g_{i}^{p}$ and
$\frac{4}{3} \epsilon \gamma^2 \vec{u}=\sum_{i=0}^{18} g_{i}
\vec{c}_{i}$. The shear viscosity is computed as
$\eta$$=$$\frac{4}{9}\gamma \epsilon(\tau - \delta t/2)c_l^2$.

\begin{figure}
  \centering
  \includegraphics[scale=0.47]{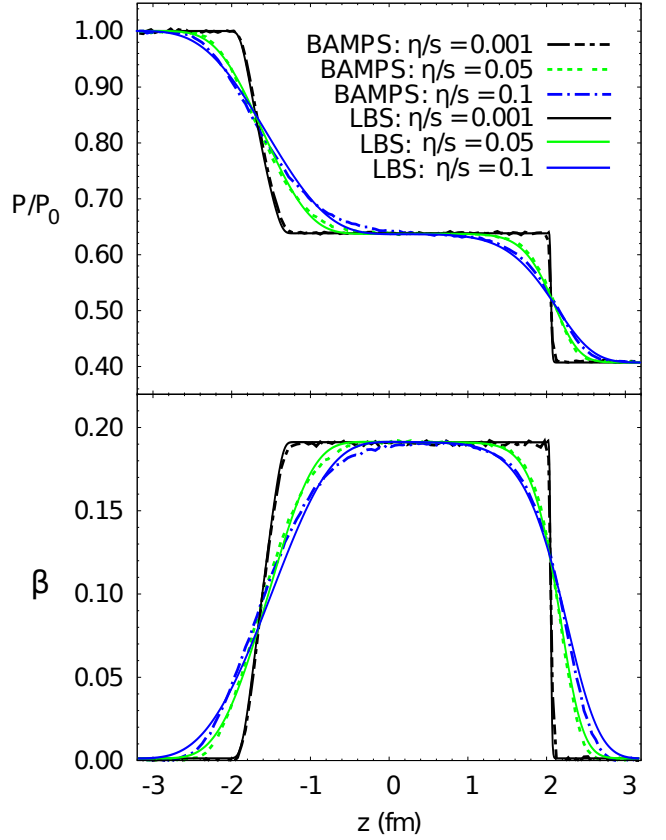}
  \caption{Comparison between the BAMPS simulations\cite{BAMPSs} and
    the lattice Boltzmann results at $t$$=$$3.2$fm/c. Note that, in
    both simulations, the value of $\beta$$\sim$$0.2$ for the speed of
    propagation of the shock wave is obtained. Pressure (top) and
    velocity (bottom) of the fluid as function of the spatial
    coordinate $z$.}\label{compare1}
\end{figure}
\begin{figure}
  \centering
  \includegraphics[scale=0.47]{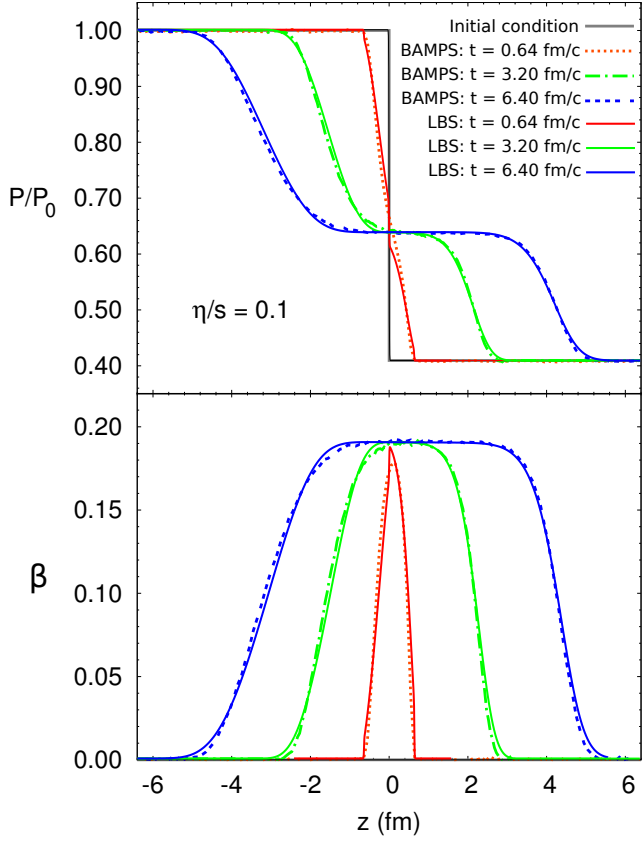}
  \caption{Time evolution of the shock wave for BAMPS
    simulations\cite{BAMPSs} and Lattice Boltzmann results. Here the
    speed of propagation of the shock wave $\beta \sim 0.2$ is
    obtained.}\label{comparetime}
\end{figure}

\begin{figure}
  \centering
  \includegraphics[scale=0.45]{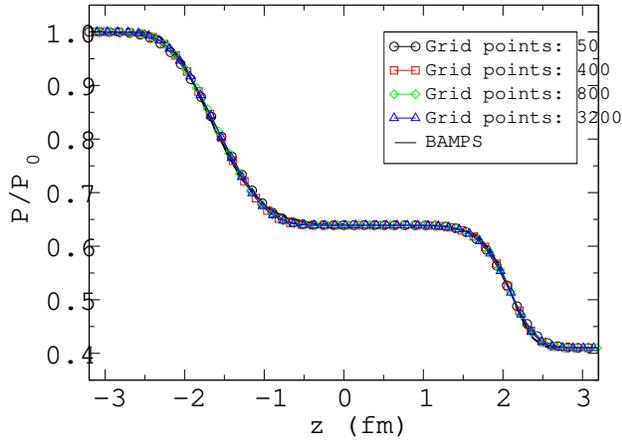}
  \caption{Lattice Boltzmann simulation of the shock wave for
    different grid resolutions, $\beta$$\sim$$0.2$ and
    $\eta/s$$=$$0.05$.}\label{convergence}
\end{figure}
\begin{figure}
  \centering
  \includegraphics[scale=0.35]{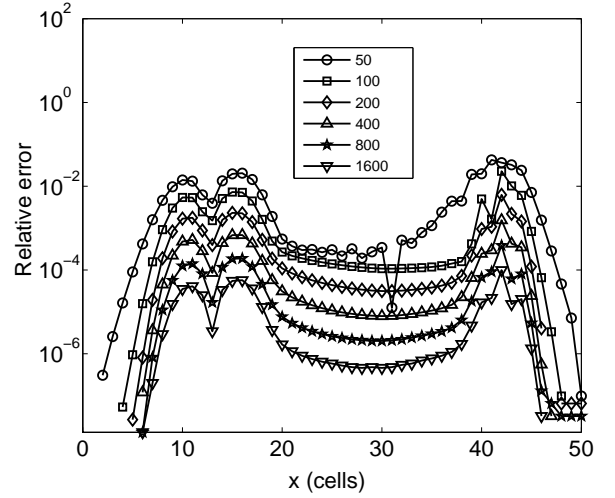}
  \caption{Relative convergence error for different grid resolutions
    as a function of the $x$-coordinate for $\beta$$\sim$$0.2$ and
    $\eta/s$$=$$0.01$.}\label{convergenceerrorvsx}
\end{figure}

\begin{figure}
  \centering
  \includegraphics[scale=0.35]{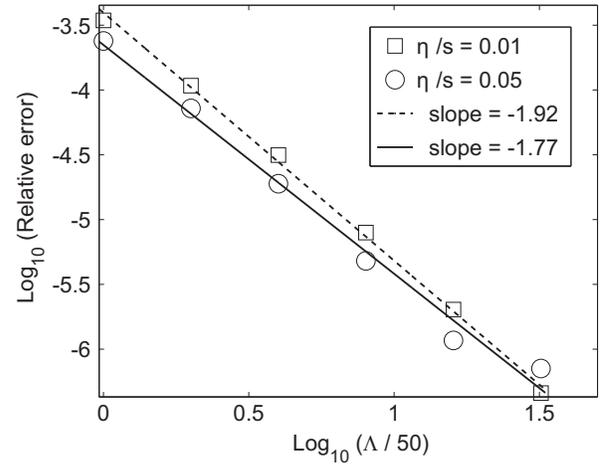}
  \caption{Relative convergence error as a function of the number of
    grid points, for $\beta$$\sim$$0.2$. $\Lambda$ represents the
    number of grid points. Here, the relative error $E_{ave}$ is
    calculated by taken the mean value of the relative errors at every
    location $z$, $E_{ave}$$=$$\frac{1}{\Lambda}\sum_{z=1}^{\Lambda}
    E_r(z)$.  }\label{convergenceerror}
\end{figure}

Also, it is worth noting that our scheme smoothly recovers the
non-relativistic limit by simply letting $\beta \rightarrow 0$.

\section{Dissipative hydrodynamics}

According to kinetic theory, dissipative effects emerge at the level
of first order terms in the Knudsen number expansion of the kinetic
equations.  At a more fundamental level, dissipation is an emergent
property resulting from the finite-time relaxation of non-equilibrium
kinetic excitations on top of the hydrodynamic "ground state".  A
detailed Chapman-Enskog analysis (see Appendix \ref{ChapmanEnskog}),
shows that the lattice formulation needs to retain second order terms
in the lattice spacing, which means that the streaming operator needs
to be expanded to second order in the lattice time step $\delta t$
and, by the light-cone rule, in $\delta x$ too.  Straightforward but
lengthy algebra, leads to the following expression of the LB dynamic
viscosity \cite{succibook}
\begin{equation}
  \label{ETA}
  \eta = \rho c_s^2 \left(\tau - \frac{\delta t}{2} \right) \equiv \rho c_s^2 \tau \left(1-\frac{\theta}{2} \right) \quad ,
\end{equation}
where we have defined
\begin{equation}
  \label{ETA}
  \theta = \frac{\delta t}{\tau} \quad ,
\end{equation}
as the parameter measuring the time-granularity of the LB scheme.
Indeed, the limit $\theta \rightarrow 0$ reproduces the continuum
value $\eta = \rho c_s^2 \tau$. Similar calculations for the
relativistic case yield (see more details in Appendix
\ref{ChapmanEnskog})

\begin{equation}
  \eta=\gamma c_s^2 \tau (\epsilon + P)(1 - \theta/2) c_l^2.  
\end{equation}

A few comments are in order. First, we note that positivity of the
kinematic viscosity implies
\begin{equation}
  0< \theta < 2 
\end{equation}
This linear stability constraint for the discrete scheme, is readily
seen to associate with the second-principle (negative viscosity
implies physical instability).

The above expressions seem to suggest that ideal hydrodynamics, i.e.
strictly zero dissipation, could be achieved in the limit $\delta t
\rightarrow 2 \tau$, i.e. $\theta \rightarrow 2$.  Actual practice,
though, shows that this limit is an illusory one, since, whenever the
viscosity falls below a given (flow-dependent) threshold, the
stability of the scheme is compromised.  Physically, the reason is
that below a given threshold, the system is no longer capable of
dissipating short-scale gradients, thereby allowing the
non-equilibrium component of the distribution function to grow wildly,
and finally ruin the simulation. This is in line with the so-called
``numerical uncertainty principle'' (NUP) for transport advection
equations, according to which a minimum non-zero viscosity is required
to secure the positivity of the positive-definite quantities, such as
the fluid density \cite{NUP}. In a nutshell, the point is that, in
order to reach zero viscosity with a positive definite distribution,
wavelengths at {\it all} scales are needed, including those below the
lattice spacing $\delta x$.  Since -by construction- the latter are
missing from a discrete lattice representation, positivity can only be
maintained through a finite amount of dissipation, typically of the
order of the inherent lattice viscosity $\frac{\delta x^2}{\delta t}$.
Incidentally, we note that viscosity has the same physical dimension
as $\delta x \delta v \sim \hbar/m$, whence the notion of
``uncertainty principle''.

For LB equations, the NUP can be formulated in terms of an inequality
involving the equilibrium and non-equilibrium components of the
discrete distribution function. To appreciate this point, let us first
recast the standard LB in the following collide-stream form:
\begin{equation}
\label{LBESC}
f_i(\vec{x} + \vec{c}_i \delta t;t+\delta t) = f'_i(\vec{x};t)
\equiv (1-\theta) f_i(\vec{x};t)+\theta f_i^{eq}
\end{equation}
where $f'$ denotes the so-called post-collisional distribution
function.

From the above, it is seen that positivity of the post-collisional
distribution at time $t$ guarantees positivity of the distribution at
the subsequent time $t+\delta t$.  Simple algebra yields:
$$
\theta < \theta_{NUP}[f] \equiv \min_i \left \lbrace
  \frac{|f_i^{eq}|}{|f_i^{neq}|} \right \rbrace
$$
This informative expression suggests the definition of three distinct
non-equilibrium regimes:
\begin{enumerate}
\item Weak non-equilibrium ($\theta_{NUP} > 2$)
\item Strong non-equilibrium ($1 < \theta_{NUP} < 2$)
\item Extreme non-equilibrium ($\theta_{NUP} < 1$)
\end{enumerate}
In the weak non-equilibrium regime (often referred to as
strong-coupling regime), the one relevant to hydrodynamics, the NUP
does not set any additional constraint to linear stability.  In the
strong non-equilibrium regime, however, non-linear stability may in
principle set the most stringent constraint.  Clearly, this is even
more so in the extreme non-equilibrium region, where the
non-equilibrium component exceeds the equilibrium one, in total
defiance of hydrodynamics.

Remarkably, LB proves capable of stable operation in this
``linearly-forbidden'' region.  In fact, the negative shift $-\delta
t/2$, (``propagation viscosity'' in LB jargon) which stems directly
from the light-cone structure of the LB streaming operator, permits to
attain very small viscosities, of order, say, $10^{-3}$ in lattice
units, while still keeping $\delta t = 1$, and $\theta \sim
2-O(10^{-3})$.  This allows for the simulation of very-low viscous
flows (such as the quark-gluon plasma) with time-steps of order
$O(1)$, which proves very beneficial for computational purposes.

The ultimate reason for such favorable behavior in the strong
non-equilibrium regime can be traced to the existence of lattice
versions of the H-theorem \cite{EntropicLB,ELB1}.

Another remarkable property of the LB formulation is that, in contrast
to hydrodynamic formulations, dissipation is not represented
explicitly through second-order spatial derivatives, but emerges
instead from a first-order, covariant {\it propagation-relaxation}
dynamics, through adiabatic enslaving of the momentum-flux tensor to
its equilibrium (ideal-hydrodynamic) expression.  As a result of this
first-order dynamics, the CFL (Courant-Friedrichs-Lewy) stability
condition of the LB scheme reads simply as $u \delta t \le \delta x$,
instead of $\nu \delta t < \delta x^2$, the latter being much more
demanding on the time-step $\delta t$, as the grid is refined ($\delta
x \rightarrow 0$).  In the above, $\nu = \frac{\mu}{\rho}$ is the
fluid kinematic viscosity.

Also to be noted, built-in causality is secured by the hyperbolic
structure of the underlying kinetic theory. 

Before closing this section, we wish to emphasize that the structure
of the dissipative terms could be enriched by turning to a multi-time
relaxation version of the collision operator, whereby different
moments relax with different rates to their equilibrium expression
\cite{HSB,MTR}.  This allows to enlarge the list of transport
coefficients, including bulk viscosity, thermal conductivity and
anisotropic transport parameters.

\section{Validation and applications}
Having discussed the basic aspects of the relativistic Lattice
Boltzmann theory, we next move on to its numerical validation and
application to two different problems of modern relativistic
hydrodynamics, namely shock propagation in viscous quark-gluon plasmas
and blast-waves from supernova explosions in interstellar media.

\subsection{Quark-Gluon Plasma}
To test the model, we solve the Riemann problem in viscous gluon
matter\cite{BAMPSs} with a ultra-relativistic equation of state
$\epsilon$$=$$3P$, as before, and the relation between energy density
and particle number density, $\epsilon$$=$$3nT$, $T$ being the
temperature\cite{RelaBoltEqua}. The initial configuration consists of
two regions, divided by a membrane located at $z$$=$$0$.  Both regions
are thermodynamically equilibrated, at different constant pressure,
$P_0$ for $z$$<$$0$ and $P_1$ for $z$$>$$0$.  At $t$$=$$0$, the
membrane is removed and the fluid starts expanding.

We implement a one-dimensional simulation with an array of size
$1$$\times$$1$$\times$$800$ using open boundary conditions at the two
ends of this $1$D chain. In this case, the $4$-velocity is given by
$u^\mu$$=$$(\gamma,0,0,\gamma\beta)^{\mu}$. The velocity of the
lattice is chosen $c_l$$=$$1.0$, so that the cell size $\delta x$ and
time step $\delta t$ are both fixed to unity.  This corresponds in IS
units to $\delta x$$=$$0.008$fm and $\delta t$$=$$0.008$fm/c.  The
viscosity is calculated as $\eta$$=$$\frac{4}{9}\gamma \epsilon(\tau -
1/2)$, and the entropy density by the approximation $s$$=$$4n-n\ln
\lambda$, with $\lambda$$=$$\frac{n}{n^{eq}}$ the gluon fugacity. The
equilibrium particle density $n^{eq}$ is given by,
$n^{eq}$$=$$\frac{d_G T^3}{\pi^2}$ with $d_G$$=$$16$ for gluons.
Next, we calculate the ratio between the viscosity and entropy
density, $\eta/s$, that is used as a parameter to characterize the
conditions for the onset of shock-waves. The pressures were chosen as
$P_0$$=$$5.43$GeVfm$^{-3}$ and $P_1$$=$$2.22$GeVfm$^{-3}$,
corresponding to $7.9433$$\times$$10^{-6}$ and
$3.2567$$\times$$10^{-6}$ in numerical units, respectively.  The
initial temperature is $T_0$$=$$350$MeV, corresponding to
$T_0$$=$$0.0287$ in numerical units. With these parameters, the
conversion between physical and numerical units for the energy, is
$1$MeV$=$$8.2$$\times$$10^{-5}$.

Fig. \ref{compare1} shows the results for different values of $\eta/s$
and the comparison with the BAMPS\cite{BAMPS} (Boltzmann Approach of
Multiparton Scattering) microscopic transport model
simulations\cite{BAMPSs} at time $3.2$fm/c.  Fig.  \ref{comparetime},
shows the time evolution of the system for $\eta/s$$=$$0.1$ for the
two numerical models.  In both cases, excellent agreement with BAMPS
is observed.  Fluids moving at higher speed, $\beta$$\sim$$0.6$, were
also considered in Ref.~\cite{rlbPRL}, where numerical ``tachyons''
with $c_l$$=$$10$ were used.

Indeed, from Eqs. \eqref{equil1} and \eqref{equil2a}, we see that the
positivity condition, $f_i^{\rm eq}$$>$$0$, implies
$\vec{c}_i\cdot\vec{u}$$<$$\frac{c_l^2}{3}$.  As a result, by raising
$c_l$, e.g. by reducing the time-step accordingly, positivity can be
preserved for higher values of $\beta$.

To check the convergence of the model, we implement simulations taking
$\eta/s$$=$$0.05$ and $\eta/s$$=$$0.01$ for different grid
resolutions. Fig.~\ref{convergence} reports the pressure profile at
time $3.2$fm/c and shows very small differences between the results
when the resolution is changed from $50$ to $3200$ grid points with
$\eta/s$$=$$0.05$. To obtain a more quantitative measure of the
convergence we use the Richardson extrapolation method
\cite{rich1,rich2}. In this method, given any quantity $A(\delta x)$
that depends on a size step $\delta x$, we can make an estimation of
order $n$ of the exact solution $A$ by using
\begin{equation}\label{richardson1}
  A = \lim_{\delta x \rightarrow 0} A(\delta x) \approx \frac{2^n A\left (\frac{\delta x}{2}\right ) - A(\delta x)}{2^n - 1} + O(\delta x^{n+1})\quad ,
\end{equation}
with errors $O(\delta x^{n+1})$ of order $n+1$. Thus the relative
error between the value $A(\delta x)$ and the ``exact'' solution $A$
can be calculated by
\begin{equation}\label{richardson2}
  E_r(\delta x) = \left |\frac{A(\delta x) - A}{A} \right | \quad .
\end{equation}
In our case, the quantity $A$ is the pressure $P(z)$ and we set up
$n=2$. We can estimate the relative error as shown in
Fig.~\ref{convergenceerrorvsx} for $\eta/s$$=$$0.01$ using
Eqs.~\eqref{richardson1} and ~\eqref{richardson2}, at every grid
point.  Indeed, the relative error with respect to the ``exact
solution'' decreases rapidly with increasing grid resolution.  More
precisely, Fig.~\ref{convergenceerror} shows that the present scheme
exhibits a near second-order convergence. This is basically in line
with the convergence properties of non-relativistic LB schemes.

However, we can see that for higher viscosity, i.e. larger values of
the relaxation time $\tau$, and higher grid resolution (smaller
$\delta x$), the order of convergence decreases due to the lack of
adiabaticity associated with increasing Knudsen number.
Nevertheless, the model is still able to reproduce shock waves, at low
resolution, hence with a very modest computational time.  For
instance, using a resolution of $50$ grid points, the simulation took
$0.94$ms in a standard PC.  Other values are shown in Table
\ref{speedtable}.
\begin{table}
  \centering
  \begin{tabular}{|c|c|c|}\hline
    Grid points & Total time steps & CPU time (ms) \\ \hline
    50 & 25 & 0.94  \\ 
    100 & 50 & 3.5 \\ 
    200 & 100 & 17.1  \\ 
    400 & 200 & 68.4  \\ 
    800 & 400 & 272  \\ 
    1600 & 800 & 1095 \\ 
    3200 & 1600 & 4396 \\ \hline
  \end{tabular}
  \caption{Computational time required for the simulation of the shock waves in quark-gluon plasma as a function of the grid resolution.}
  \label{speedtable}  
\end{table}
From this table, it is readily appreciated that the computational cost scales
linearly with the number of grid points and time-steps.

\subsection{Supernova explosion simulation}

\begin{figure}        
  \centering \subfigure[]{
    \includegraphics[trim = 25mm 13mm 25mm 12mm, clip, width=7.7cm,
    height=4.5cm]{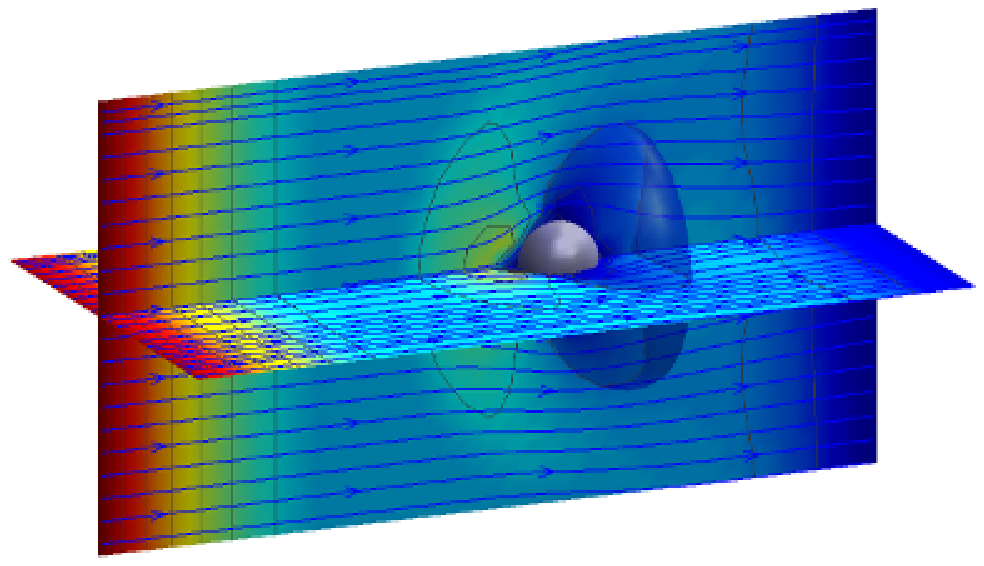}
    \label{3Da}
  } \centering \subfigure[]{
    \includegraphics[trim = 25mm 13mm 25mm 12mm, clip, width=7.7cm,
    height=4.5cm]{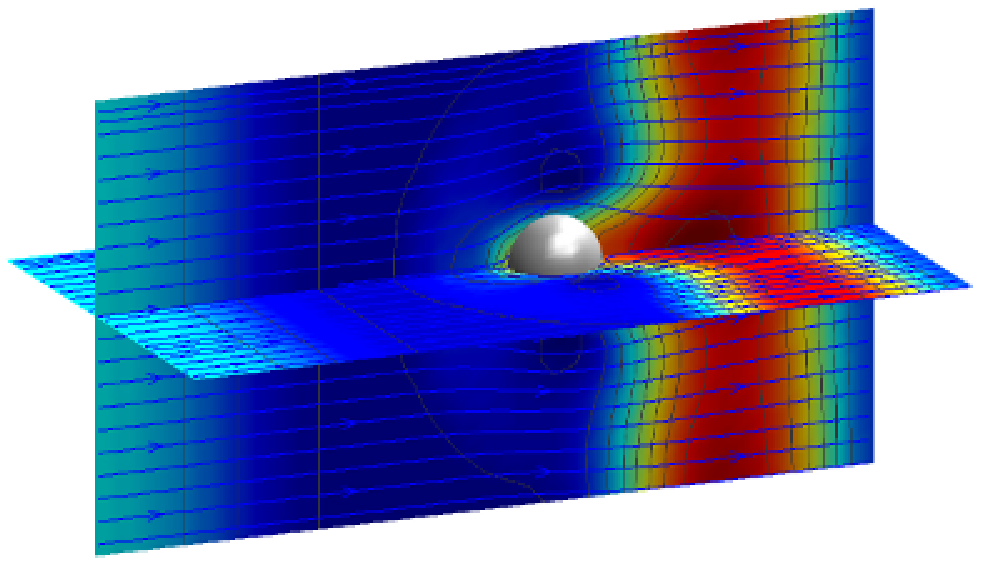}
    \label{3Db}
  } \centering \subfigure[]{
    \includegraphics[trim = 25mm 13mm 25mm 12mm, clip,
    width=7.7cm, height=4.5cm]{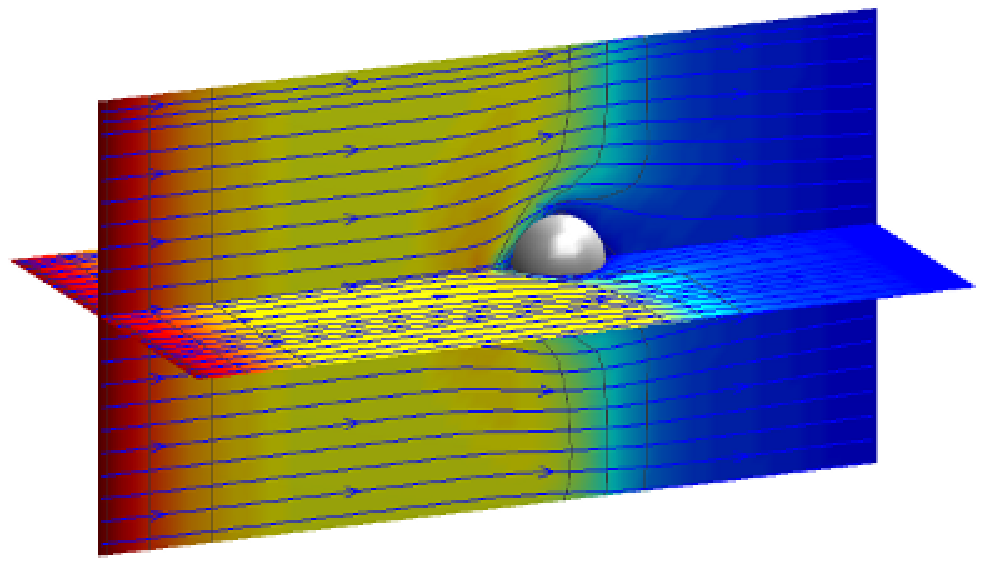}
    \label{3Dc}
  }
  \caption{\label{picture3D} Relativistic shock wave, generated by a
    $\gamma$-ray burst or $X$-ray flash supernova explosion
    \cite{supernova2, supernova3}, impacting on a massive interstellar
    cloud at $|\vec{\beta}|=0.5$ at $t=1350$ time steps which is
    equivalent to $4280$ years. Here the streamlines represent the
    velocity field, and the colors \subref{3Da} the pressure,
    \subref{3Db} the particles density, and \subref{3Dc} the
    temperature. The simulation was implemented on a grid of $200
    \times 100 \times 100$ cells.}
\end{figure}

Several important astrophysical phenomena involve
strongly-relativistic hydrodynamics, and some of them fall in the
region of $\gamma \sim 1.4$, covered by our scheme.  This is the case,
for instance, of blastwaves produced by supernova explosions
\cite{supernova3}.  In this section, we simulate a shock wave,
generated by, say, a GRB ($\gamma$-ray burst) or XRF ($X$-ray flash)
supernova explosion \cite{supernova2, supernova3}, colliding against
an interstellar cloud composed by massive matter, e.g. molecular
gas\cite{Scipaper}.  The ejecta from the explosion of such supernovae
are known to sweep the interstellar material along, up to relativistic
velocities (relativistic outflows)
\cite{supernova1,supernova2,supernova3}.

The simulation is implemented in a box of size $6\times 3\times 3$
$\times 10^{16}$ Km in a coordinate system $(x,y,z)$, using a lattice
of $200 \times 100 \times 100$ cells, which gives a cell length
$\delta x$$=$$\delta y$$=$$\delta z$$=$$3\times 10^{14}$ Km, using
numerical ``tachyons'' with $c_l$$=$$10$, and a time step $\delta
t$$=$$3.17$ years. The simulation region is divided in two zones by
the plane $x=50$. The interstellar medium, located at $x>50$, is
characterized by a particle density $n_1$$=$$0.6$ cm$^{-3}$ and
temperature $T_1$$=$$10^4$ K. The massive cloud is modeled as a
spherical obstacle, with a radius of $10$ cells, centered at location
$(100,50,50)$. The boundary condition on the surface of the obstacle
is implemented forcing the obstacle cells to evolve to the equilibrium
distribution function with the constant values, $n=n_1$, $\vec{u}=0$,
and $T=T_1$. Open boundary condition was implemented at right, left,
top, bottom and front of the simulation zone according to the shock
wave propagation direction ($x$-direction), which consists on copying
the information of the distribution functions from the second last
cells to the last ones of the boundary.  At back boundary we set an
inlet flow boundary condition fixing the distribution functions of the
boundary cells with the equilibrium distribution function evaluated
with the initial conditions $n_0$ and $T_0$ \cite{succibook,benzi1}.
In order to obtain a shock wave moving at $|\vec{\beta}| \simeq 0.5$
along the $x$-direction, we set $T_0$$=$$6 T_1$ and $n_0$$=$$2 n_1$
for the region $x\leq 50$.  The simulation, spanning $1350$ time
steps, takes about $1900$ CPU seconds on a standard PC.
Fig.~\ref{picture3D} shows the simulation results for the velocity,
pressure, particle density, and temperature fields of the supernova
remnant, during the impact of the shock wave on the massive
interstellar cloud, red and blue denoting high and low values,
respectively.
\begin{figure}
  \centering
  \includegraphics[scale=0.75]{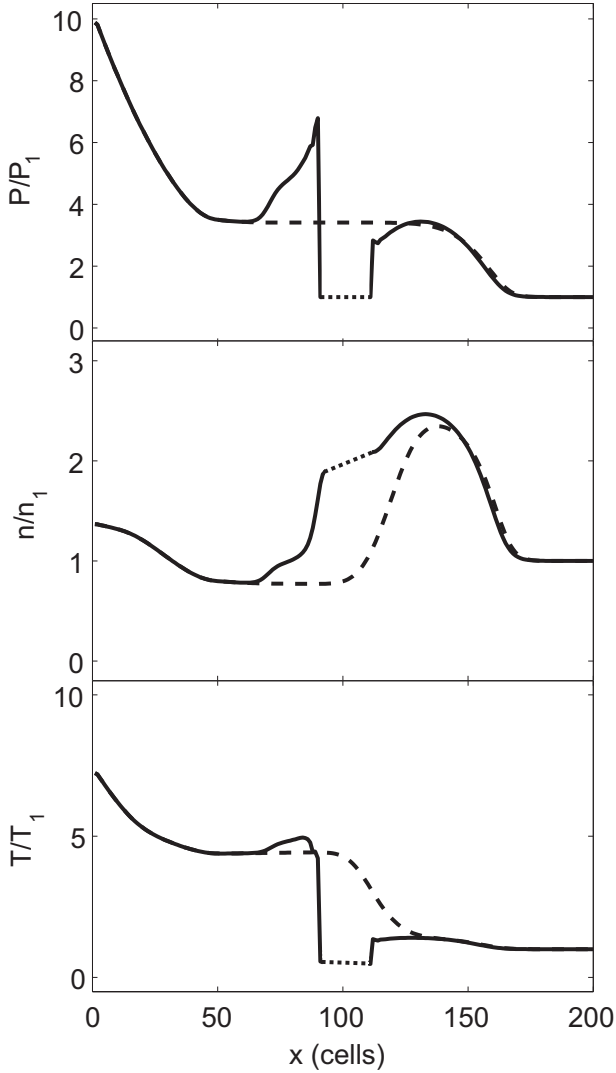}
  \caption{Pressure $P$, number of particles density $n$, and
    temperature $T$ of the supernova remnant as a function of the $x$
    coordinate at $y=z=50$ and $t=1350$ time steps equivalent to
    $4280$ years. The solid lines describe the values in the presence
    of the obstacle, dotted line the region where the massive
    interstellar cloud is located, and the dashed line the values
    without obstacle.  }\label{numerical}
\end{figure}

Here, we can see that the density $n$ is higher in the shock front,
due to sweeping of interstellar material by the shock-wave, which is
compressing the fluid.  On the other hand, the temperature of the
fluid is higher in the zone of $x\leq 50$, as a consequence of the
initial configuration. The temperature is seen to increase in the zone
where the collision takes place (see Fig.~\ref{numerical}), and so
does the temperature. This is due to conversion of kinetic energy to
pressure/temperature caused by the momentum lost on the solid boundary
of the massive cloud. 

\begin{figure}     
  \centering
  \subfigure[]{
    \includegraphics[trim = 30mm 0mm 20mm 0mm, clip, scale=0.8]{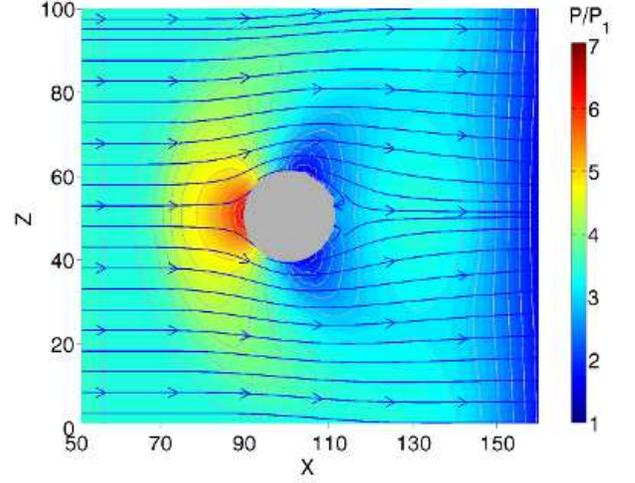}
    \label{3D2a}
  }
  \subfigure[]{
    \includegraphics[trim = 30mm 0mm 20mm 0mm, clip, scale=0.8]{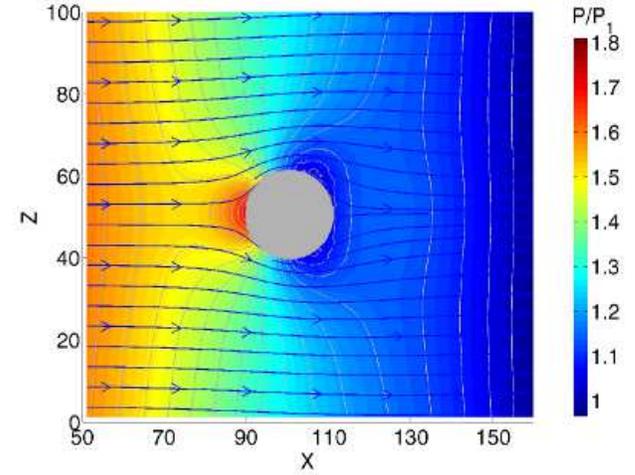}
    \label{3D2b}
  }
  \subfigure[]{
    \includegraphics[trim = 30mm 0mm 20mm 0mm, clip, scale=0.8]{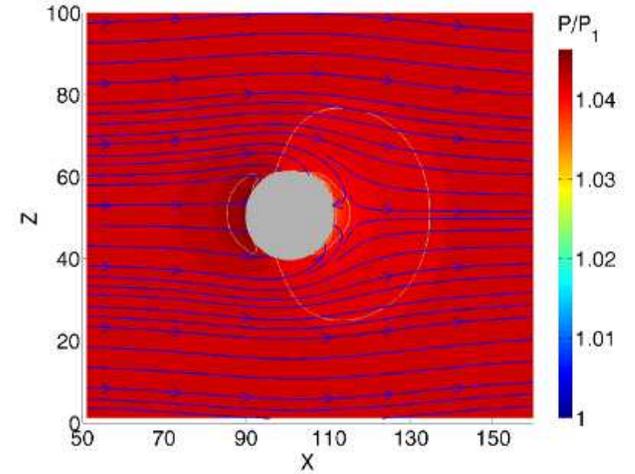}
    \label{3D2c}
  }
  \caption{\label{figpressure} Fluid pressure after the collision of
    the shock wave, produced by a supernova explosion, against the
    massive interstellar cloud, at \subref{3D2a} $|\vec{\beta}|=0.5$,
    \subref{3D2b} $|\vec{\beta}|=0.2$, and \subref{3D2c}
    $|\vec{\beta}|=0.01$ in a cut going through the center of the
    cloud.  The streamlines represent the velocity field, and the
    colors the pressure.}
\end{figure}

\begin{figure}  
  \centering
  \subfigure[]{
    \includegraphics[trim = 30mm 0mm 20mm 0mm, clip, scale=0.8]{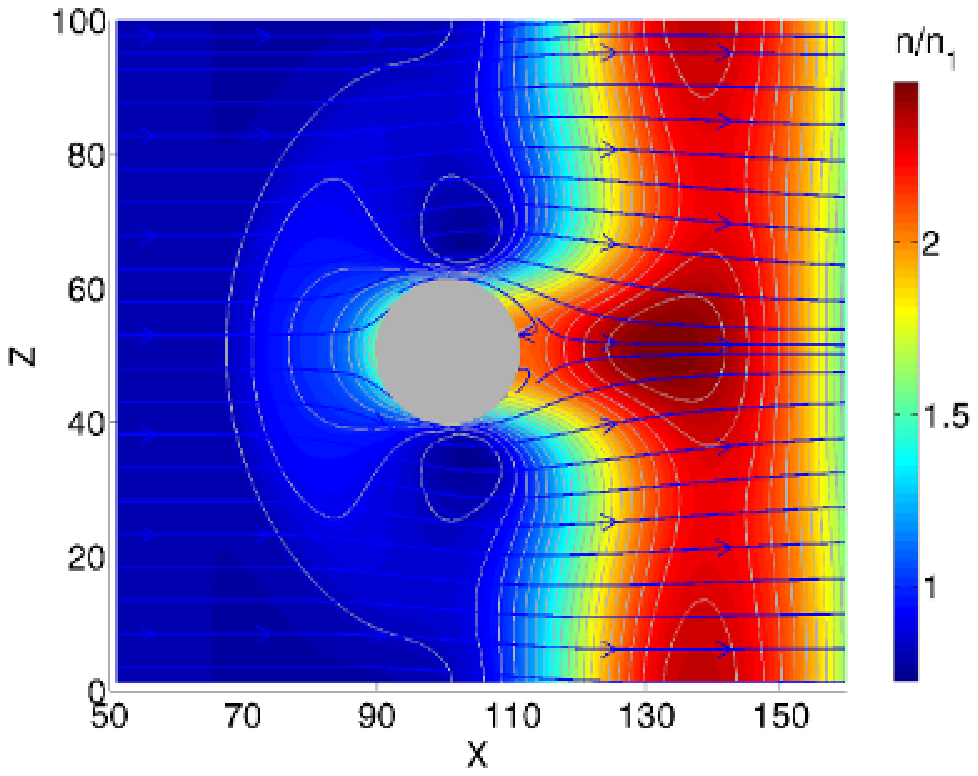}
    \label{3D3a}
  }
  \subfigure[]{
    \includegraphics[trim = 30mm 0mm 20mm 0mm, clip, scale=0.8]{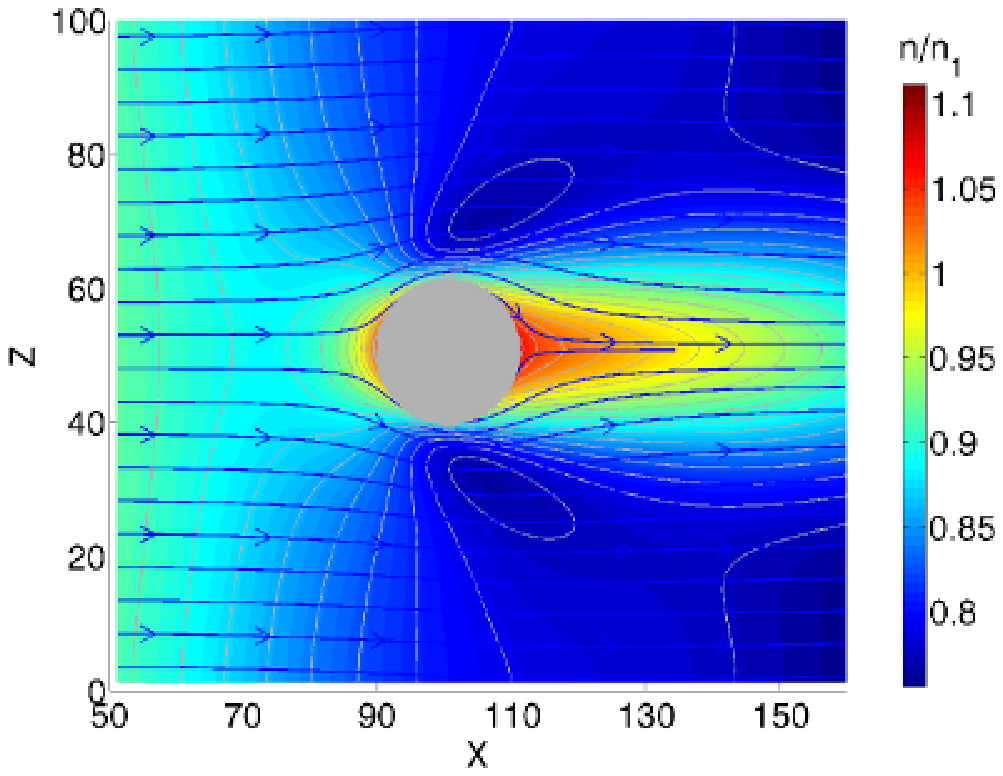}
    \label{3D3b}
  }
  \subfigure[]{
    \includegraphics[trim = 30mm 0mm 20mm 0mm, clip, scale=0.8]{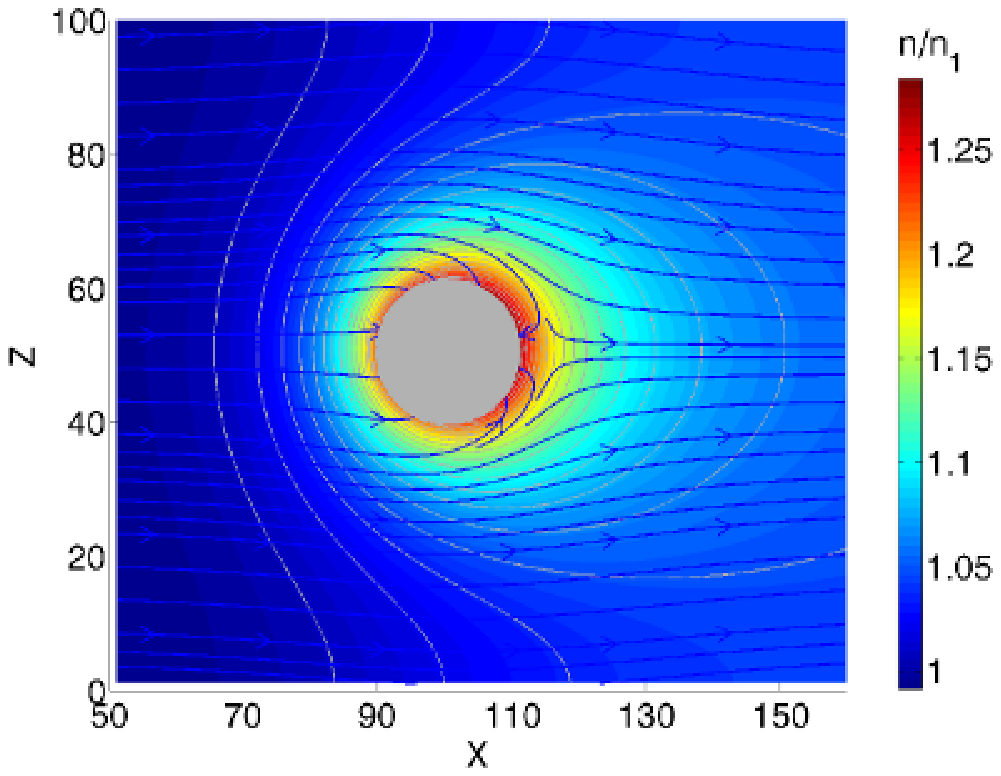}
    \label{3D3c}
  }
  \caption{\label{figdensidad} Particles density of the fluid after
    the collision of the shock wave, produced by a supernova
    explosion, against the massive interstellar cloud, at
    \subref{3D2a} $|\vec{\beta}|=0.5$, \subref{3D2b}
    $|\vec{\beta}|=0.2$, and \subref{3D2c} $|\vec{\beta}|=0.01$ in a
    cut going through the center of the cloud.  The streamlines
    represent the velocity field, and the colors the particles
    density.}
\end{figure}

Fig.~\ref{numerical} illustrates in more detail the density $n$,
pressure $P$ and temperature $T$ of the fluid during the collision and
compares the respective curves with the ones obtained when the
obstacle is absent. Note that the particle density, pressure, and
temperature values, with and without obstacle, present a small
difference sufficiently downstream the obstacle along the $x$-axis at
$y=z=50$ (see Fig.~\ref{numerical}). During the collision, the
shockwave surrounds the obstacle and later the fluid meets again at
the $x$-axis and overlaps. Due to this, the $x$-component of the
shockwave propagation velocities are the same (because of symmetry)
for all the incoming fluid to the meeting zone, the perturbations
along this axis close to the shock front are weak, contrary to the zone
near the obstacle, where the fluid fills up again, due to the low pressure.
However, the fluid moves slower than in the case without obstacle
because of the existence of flow moving outwards off the axis. 
If we increase the ratio between the cross section and the
length of the obstacle, larger departures between the velocity of the
shock-fronts with and without obstacle would be expected. 
Moreover, later in time after the collision, differences in the pressure
and other quantities, can generate turbulence. 
Transversal perturbations in the variables, as one moves out from the $x$-axis, are
shown in Figs.~\ref{picture3D}, \ref{figpressure} and \ref{figdensidad}.

Shock waves form when the speed of injection of mass exceeds the sound
speed of the surrounding medium \cite{Scipaper}. By changing the
values of the temperature of the fluid in the region $x\leq 50$, in
order to obtain speeds of mass injection of $|\vec{\beta}|=0.5$,
$|\vec{\beta}|=0.2$, and $|\vec{\beta}|=0.01$, we can see that the
increment of the particle density due to the sweeping of interstellar
medium by the shock wave becomes appreciable only for
$|\vec{\beta}|=0.5$ (see Fig.~\ref{figdensidad}).  A similar argument
applies to the pressure cone (see
Fig.~\ref{figpressure}\subref{3D2a}).  Indeed, in the other cases, the
speed of mass injection is lower than the sound speed, and therefore
no shock-wave can be formed.

\section{Conclusions and outlook}

In this paper, we have provided a detailed discussion of the Lattice
Boltzmann formulation for relativistic fluids.  In particular, details
on the construction of the relevant lattice equilibria are provided,
emphasizing the common aspects with standard Lattice Boltzmann theory.

The scheme is shown to exhibit excellent agreement with previous
numerical simulations of shock wave propagation in quark-gluon
plasmas, at a fraction of the cost of hydrodynamic codes.  
Near-second order accuracy with grid resolution and linear
computational time with space-time resolution, are evidenced.

As an example of relativistic hydrodynamics with non-trivial
geometries, we have also applied our scheme to an astrophysical
system, namely the collision of a shock wave, produced by a supernova
explosion, against a cold molecular cloud.  The numerical simulations
show good qualitative results yielding information, that can be
compared with experimental results and other numerical methods.

For the case of quark-gluon plasma simulations, the present
lattice-kinetic algorithm appears to be nearly an order of magnitude
faster than corresponding hydrodynamic codes.  This is due to the fact
that, at variance with any hydrodynamic representation, LB moves
information along constant light-cones rather than space-time changing
material fluid streamlines \cite{Paulcomment}.  This trivializes the
Riemann problem to a mere shift of the distribution function along the
corresponding lightcone, a {\it floating-point free}, {\it exact}
operation, which is way more convenient than propagating hydrodynamic
fields along space-time changing streamlines.  Such an advantage, key
in ordinary lattice Boltzmann fluids, might be even accrued in
the relativistic context.

Several issues remain open for future research. First, extensions of
the present scheme to higher-order lattices are worth being
considered, for they should give access to higher values of $\beta$,
by use of correspondingly higher-order lattice equilibria.  This
strategy has indeed proved very effective for the case of compressible
and thermal non-relativistic fluids
\cite{compress1,compress2,compress3,AIGUOEPL}.

Another important question concerns the existence of a relativistic
lattice H-theorem.  Apart from the theoretical interest on its own,
this has major implications on the numerical stability of the scheme
at high Reynolds number, i.e. for the simulation of relativistic
turbulence \cite{Turbu1}.

Yet another interesting research direction is the simulation of
relativistic flows with a non-ideal equation of state, which may find
applications in relativistic cosmology and high-energy theories of the
early universe \cite{Paul1,EoS1}.

These are just but a few of the many exciting developments and
applications which may currently be envisaged for the relativistic
Lattice Boltzmann equation presented in this paper.

\begin{acknowledgments}
  The authors are grateful to P. Romatsche for many valuable
  suggestions. SS would like to acknowledge kind hospitality and
  financial support from ETH Z\"urich, Tufts University and
  the Programme "Partial Differential Equations in Kinetic Theories" at
  the Isaac Newton Institute for Mathematical Sciences, Cambridge, UK. 
  
  BB would like to acknowledge NSF grant 0619447, and TeraGrid allocation
  MCA08X031. MM and HH are grateful for the financial support of the
  Swiss National Science Foundation (SNF) under Grant No. 116052.
\end{acknowledgments}

\appendix
\section{Moment Matching Procedure}
\label{LagrangeParameters}

To obtain the equilibrium distribution functions $f_i^{\rm eq}$ and
$g_i^{\rm eq}$ that reproduce in the continuum limit the hydrodynamic
equations, Eqs.~\eqref{macroeq10} and \eqref{macroeq20}, we use the
moment-matching procedure. In section \ref{sec:lbmodel}, we describe
the procedure and calculate the equilibrium distribution functions
$f_i^{\rm eq}$ in order to obtain the conservation of particle number,
Eq.~\eqref{macroeq20}. Following a similar procedure, to find the
equilibrium distributions $g_i^{\rm eq}$, first we can write it, as
before, as
\begin{subequations}\label{ap:equil0}
  \begin{equation}
    g_i^{\rm eq} = w_i [C + \vec{c}_i\cdot \vec{D} + \tensor{E}:(\vec{c}_i \vec{c_i} - \alpha  \tensor{I})] \quad , {\rm for} \quad i > 0 \quad ,
  \end{equation}
  \begin{equation}
    g_0^{\rm eq} = w_0 [F] \quad ,
  \end{equation}
\end{subequations}
with $C$, $\vec{D}$, $\alpha$, and $\tensor{E}$ the Lagrange
multipliers. Then, we impose the following constraints:
\begin{equation}\label{eq:0}
  \sum_{i=0}^{18} g_i^{\rm eq} = \gamma^2 (\epsilon + P) - P \quad ,
\end{equation}
\begin{equation}\label{eq:1}
  \sum_{i=0}^{18} g_i^{\rm eq} \vec{c}_i = (\epsilon + P) \gamma^2 \vec{u} \quad .
\end{equation}
and additionally,
\begin{equation}\label{eq:2}
  \sum_{i=0}^{18} g_i^{\rm eq} c_{ia} c_{i\beta} = P\delta_{a b} + (\epsilon + P)\gamma^2 u_a u_b \quad .
\end{equation}
Replacing Eq.~\eqref{ap:equil0} into Eq.~\eqref{eq:0}, \eqref{eq:1},
and \eqref{eq:2}, and summing up over the index $i$, we obtain
\begin{equation}\label{eq:01}
  \frac{1}{3}\left(2C + F + {\rm Tr}(\tensor{E})(c_l^2 -2\alpha) \right ) = \gamma^2 (\epsilon + P) - P \quad ,
\end{equation}
\begin{equation}\label{eq:11}
  \frac{c_l^2}{3}\vec{D} = (\epsilon + P) \gamma^2 \vec{u} \quad ,
\end{equation}
and
\begin{equation}\label{eq:21}
  \begin{aligned}
    \frac{c_l^2}{9}\left(3C + (c_l^2-3\alpha){\rm Tr}(\tensor{E})
    \right)\delta_{ab} &+ \frac{2c_l^4}{9}E_{a b} = P\delta_{a b} \\
    &+ (\epsilon + P)\gamma^2 u_a u_b \quad ,
  \end{aligned}
\end{equation}
where we have defined ${\rm Tr}(\tensor{E})$ as the trace of the
tensor $\tensor{E}$. From Eq.~\eqref{eq:11} we can see that
$\vec{D}$$=$$\frac{3}{c_l^2}(\epsilon + P)\gamma^2 \vec{u}$. If we
compare the left and right hand sides of Eq.~\eqref{eq:21}, we can
conclude that $\alpha$$=$$\frac{c_l^2}{3}$, and therefore
Eq.~\eqref{eq:21} is simplified to
\begin{equation}\label{eq:22}
  \begin{aligned}
    \frac{c_l^2}{3} C \delta_{ab} + \frac{2c_l^4}{9}E_{a b} =
    P\delta_{a b} + (\epsilon + P)\gamma^2 u_a u_b \quad .
  \end{aligned}
\end{equation}
Comparing again both sides of this equation the Lagrange multipliers
$C$$=$$\frac{3P}{c_l^2}$ and $E_{ab}$$=$$\frac{9}{2c_l^4}(\epsilon +
P)\gamma^2 u_a u_b$ are obtained. Now, the only missing parameter to
be determined is $F$. Replacing the values of $C$, $\alpha$, and
$\tensor{E}$ into Eq.~\eqref{eq:01}, it gives
\begin{equation}\label{eq:02}
  \frac{2P}{c_l^2} + \frac{F}{3} + \frac{c_l^2}{9}{\rm Tr}(\tensor{E}) = \gamma^2 (\epsilon + P) - P \quad .
\end{equation}
From here, we can get the Lagrange parameter $F$ and it can be written
as
\begin{equation}
  F=(\epsilon + P)\gamma^2 \left[3 -
    3\frac{(2+c_l^2)P}{c_l^2(\epsilon + P)\gamma^2} -
    \frac{3}{2c_l^2}(\epsilon + P)\gamma^2 |\vec{u}|^2 \right] \quad.
\end{equation}

Summarizing, we have determined all the Lagrange parameters and
therefore the equilibrium distribution functions $g_i^{\rm eq}$ that
recover in the continuum limit the conservation equation for the
momentum-energy.

\section{Chapman-Enskog Expansion} 
\label{ChapmanEnskog}

The discrete Boltzmann equations, Eqs.~\eqref{lbe1} and \eqref{lbe2},
determine the evolution of the lattice relativistic fluid.  
In the continuum limit, these
evolution rules must reproduce the partial differential equations of
relativistic hydrodynamics.  In order to accomplish this task, we
adopt a standard Chapman-Enskog expansion.  We start by taking the
Taylor expansion of the Boltzmann equations, up to second order in
spatial and temporal coordinates,
\begin{subequations}
  \begin{eqnarray}{\label{lbee1ED}}
    \begin{aligned}
      & v_{ia}\partial_a f_i +\frac{1}{2} \sum_{a,b}
      \partial_{a}\partial_{b} f_i v_{ia} v_{ib} +\partial_t f_i \\ &+
      \partial_t v_{ia} \partial_a f_i +
      \frac{1}{2}\partial^2_t f_i\delta t^2 =
      -\frac{1}{\tau}(f_i-f_i^{\rm eq}) \quad ,
    \end{aligned}
  \end{eqnarray}
  \begin{eqnarray}{\label{lbee2ED}}
    \begin{aligned}
      & v_{ia}\partial_a g_i +\frac{1}{2} \sum_{a,b}
      \partial_{a}\partial_{b} g_i v_{ia} v_{ib} +\partial_t g_i \\ &+
      \partial_t v_{ia} \partial_a g_i +
      \frac{1}{2}\partial^2_t g_i\delta t^2 =
      -\frac{1}{\tau}(g_i-g_i^{\rm eq}) \quad ,
    \end{aligned}
  \end{eqnarray}
\end{subequations}
where $a, b$$=$$x, y, z$ denote the $x$, $y$ and $z$ components.
Next, we expand the distribution functions, and the space-time
derivatives in a power series of the Knudsen number $\kappa$, as
follows:
\begin{subequations}\label{knudsenexpansion}
  \begin{equation}
    f_i=f_i^{(0)}+\kappa f_i^{(1)}+ \kappa^2 f_i^{(2)}+ ... \quad ,
  \end{equation}
  \begin{equation}
    g_i=g_i^{(0)}+\kappa g_i^{(1)}+ \kappa^2 g_i^{(2)}+ ... \quad ,
  \end{equation}
  \begin{equation}
    \partial_t=\kappa \partial_{t_1}+\kappa^2 \partial_{t_2} +... \quad ,
  \end{equation}
  \begin{equation}
    \partial_{a}=\kappa \partial_{1 a}+\kappa^2 \partial_{2 a}... \quad .
  \end{equation}
\end{subequations}
It is assumed that only the $0$th order terms of the distribution
functions contribute to the macroscopic conserved variables.
Therefore, for $n>0$ we have
\begin{subequations}{\label{nomacrosED}}
  \begin{equation}
    \sum_{i} f_{i}^{(n)} =0 \quad , \quad \sum_{i} g_{i}^{(n)} =0 \quad ,
  \end{equation}
  \begin{equation}
    \sum_{i} f_{i}^{(n)} \vec{v}_{i}=0 \quad , \quad \sum_{i} g_{i}^{(n)} \vec{v}_{i}=0 \quad .
  \end{equation}
\end{subequations}

By inserting these results into Eqs.\eqref{lbee1ED} and
\eqref{lbee2ED}, we obtain at $0$th-order in $\kappa$
\begin{equation}{\label{zeroth1}}
  f_{i}^{\rm eq}=f_{i}^{(0)} \quad ,
\end{equation}
\begin{equation}{\label{zeroth2}}
  g_{i}^{\rm eq}=g_{i}^{(0)} \quad ,
\end{equation}
to the first order in $\kappa$,
\begin{subequations}{\label{firstED}}
  \begin{eqnarray}{\label{firstbED}}
    \begin{aligned}
      v_{ia} \partial_{1a} f_{i}^{(0)} +\partial_{t_1}
      f_{i}^{(0)} = -\frac{f_{i}^{(1)}}{\tau} \quad ,
    \end{aligned}
  \end{eqnarray}
  \begin{eqnarray}{\label{firstcED}}
    \begin{aligned}
      v_{ia} \partial_{1a} g_{i}^{(0)} +\partial_{t_1}
      g_{i}^{(0)}= -\frac{g_{i}^{(1)}}{\tau} \quad ,
    \end{aligned}
  \end{eqnarray}
\end{subequations}
and to the second order in $\kappa$,
\begin{subequations}{\label{secondED}}
\begin{eqnarray}{\label{secondbED}}
  \begin{aligned}
    \biggl(1& -\frac{1}{2\tau}\biggr)\left(
      v_{ia}\partial_{1a} +\partial_{t_1}\right)f_{i}^{(1)} \\
    &+ \partial_{t_2} f_{i}^{(0)} + v_{ia}
  \partial_{2a} f_i^{(0)} =-\frac{f_{i}^{(2)}}{\tau} \quad .
  \end{aligned}
\end{eqnarray}
\begin{eqnarray}{\label{secondcED}}
  \begin{aligned}
    \biggl(1& -\frac{1}{2\tau}\biggr)\left(v_{ia}
      \partial_{1a} +\partial_{t_1}\right)g_{i}^{(1)} \\ &+
    \partial_{t_2} g_{i}^{(0)} + {v}_{ia} \partial_{2a}
    g_i^{(0)} =-\frac{g_{i}^{(2)}}{\tau} \quad .
  \end{aligned}
\end{eqnarray}
\end{subequations}

A this stage, all the ingredients required to determine the
equations that the model satisfies in the continuum limit, are
available.  By summing up Eqs.  \eqref{firstbED}, \eqref{firstcED},
\eqref{secondbED}, and \eqref{secondcED} over index $i$, taking into
account Eqs.~\eqref{zeroth1}, \eqref{zeroth2}, and the equilibrium
distribution functions defined by Eqs.~\eqref{equil1s},
\eqref{equil2as}, and \eqref{equil2bs}, we obtain
\begin{eqnarray}{\label{Max00ED1}}
  \partial_{t_1} (n\gamma) + \partial_{1a} (n \gamma u_a)=0 \quad ,
\end{eqnarray}
\begin{eqnarray}{\label{Max00ED2}}
  \partial_{t_1} ( (\epsilon + P)\gamma^2 - P ) + \partial_{1a} ( (\epsilon+P) \gamma^2 u_{a})=0 \quad ,
\end{eqnarray}
and
\begin{eqnarray}{\label{Max01ED1}}
  \partial_{t_2} (n\gamma) + \partial_{2a} (n \gamma u_a)=0 \quad .
\end{eqnarray}\
\begin{eqnarray}{\label{Max01ED2}}
  \partial_{t_2} ( (\epsilon + P)\gamma^2 - P  ) + \partial_{2a} ( (\epsilon + P) \gamma^2 u_a )=0 \quad .
\end{eqnarray}

By adding these equations, the first and second scalar equations,
associated with the conservation of the number of particle and the
first conservation equation for the momentum-energy,
\begin{eqnarray}{\label{Max02ED1}}
  \partial_t (n\gamma) + \partial_a (n \gamma u_a)=0 \quad  ,
\end{eqnarray}
and
\begin{eqnarray}{\label{Max02ED2}}
  \partial_t ( (\epsilon + P)\gamma^2 - P  ) + \partial_a ( (\epsilon + P) \gamma^2 u_{a})=0 \quad ,
\end{eqnarray}
are obtained, which correspond to Eqs.\eqref{macroeq2} and
\eqref{macroeq1a}, respectively.  To derive the second conservation
equation, Eq.\eqref{macroeq1b}, the equations \eqref{firstcED} and
\eqref{secondcED} must be multiplied by $\vec{v}_i$ and summed up over
the index $i$, which leads to
\begin{equation}\label{Max10ED}
  \begin{aligned}
    \partial_{t_1}[ (\epsilon &+P)\gamma^2 u_b ] +
    \partial_{1b}P \\&+ \partial_{1a} \left[ (\epsilon + P)\gamma^2
      u_a u_b \right] = 0\quad ,
  \end{aligned}
\end{equation}
and
\begin{equation}\label{Max11ED}
  \begin{aligned}
    \partial_{t_2} [ (\epsilon &+P)\gamma^2 u_b ] +
    \partial_{2b}P \\ &+ \partial_{2a} \left[ (\epsilon + P)\gamma^2
      u_a u_b \right] +\partial_{1a} \Pi^{(1)}_{a b} = 0 \quad ,
  \end{aligned}
\end{equation}
where the first order tensor $\Pi_{a b}^{(1)}=
\left(1-\frac{1}{2\tau}\right)\sum_{i} g_i^{(1)}v_{ia} v_{ib}$ is
defined. By replacing the distribution function $f_i^{(1)}$ from
Eq.\eqref{firstcED} into the tensor $\Pi_{a b}^{(1)}$, and the result
into Eq.\eqref{Max11ED}, we obtain
\begin{equation}\label{Max11ED2}
  \begin{aligned}
    \partial_{t_2}&[(\epsilon+P)\gamma^2 u_b ] +
    \partial_{2b}P + \partial_{2a} \left[ (\epsilon + P)\gamma^2 u_a
      u_b \right] \\ &- \partial_{1a} \left[ \partial_{1b} (\eta
      \gamma u_a)+\partial_{1a}(\eta \gamma u_b) + \partial_{1l} (\eta
      \gamma u_l)\delta_{ab} \right] = 0 \quad ,
  \end{aligned}
\end{equation}
with the viscosity $\eta$$=$$\frac{1}{3}\gamma (\epsilon + P)(\tau -
\delta t/2)c_l^2$, $l$ denoting again the spatial components. To
arrive to these results, we have assumed low-speed, $|\vec{u}|\ll c$.
The second momentum-energy conservation equation,
Eq.\eqref{macroeq1b}, is obtained by summing up Eqs.\eqref{Max11ED2}
and \eqref{Max10ED}. It gives
\begin{equation}\label{Max11ED3}
  \begin{aligned}
    \partial_{t}&[(\epsilon+P)\gamma^2 u_b ] +
    \partial_{b}P + \partial_{a} \left[ (\epsilon + P)\gamma^2 u_a u_b
    \right] \\ &- \partial_{a} \left[ \partial_{b} (\eta \gamma
      u_a)+\partial_{a}(\eta \gamma u_b) + \partial_{l} (\eta \gamma
      u_l)\delta_{ab} \right] = 0 \quad .
  \end{aligned}
\end{equation}
The derivation of the dissipative term associated with the viscosity
$\eta$, in Eq.~\eqref{Max11ED3}, is obtained assuming low values of
$\beta$ to neglect higher order terms ($\sim |\vec{u}|^3$)
contributions.

Summarizing, Eqs. \eqref{Max02ED1}, \eqref{Max02ED2} and
\eqref{Max11ED3} determine the evolution of the fluid, according to
the relativistic hydrodynamics equations.

\bibliography{PRD}

\end{document}